\documentclass[fleqn,usenatbib]{mnras}

\usepackage{newtxtext,newtxmath}
\usepackage[T1]{fontenc}

\DeclareRobustCommand{\VAN}[3]{#2}
\let\VANthebibliography\thebibliography
\def\thebibliography{\DeclareRobustCommand{\VAN}[3]{##3}\VANthebibliography}

\usepackage{graphicx}	
\usepackage{amsmath}	
\usepackage{hyperref}

\newcommand{\kmsmpc}{\kms\;{\rm Mpc}^{-1}}

\newcommand{\hmpc}{h^{-1}{\rm Mpc}}

\newcommand{\kms}{\;{\rm km}\,{\rm s}^{-1}}

\newcommand{\simba}{{\sc Simba}}

\newcommand{\fgas}{f_{\rm gas}}
\newcommand{\fedd}{f_{\rm Edd}}
\newcommand{\mbh}{\;{M}_{\rm BH}}
\newcommand{\mstar}{\;{M}_{*}}

\newcommand{\ro}{R_{0}}
\newcommand{\msun}{\;{\rm M}_{\odot}}

\newcommand{\mdot}{\;{\dot{M}}_{\rm BH}}

\newcommand{\WHz}{\rm{~W~Hz}^{-1}}
\newcommand{\power}{\;P_{\rm 1.4~GHz}}

\newcommand{\lbol}{L_{\rm bol}}
\newcommand{\lmech}{L_{\rm mech}}

\newcommand{\ledd}{L_{\rm Edd}}

\title[MIGHTEE \simba]{Radio Galaxies in SIMBA: A MIGHTEE Comparison}

\author[N.L. Thomas et al. ]{Nicole~L. Thomas,$^{1,2,3}$\thanks{E-mail: thomas.nicolelynn@gmail.com}\,
Imogen~H. Whittam,$^{4,5}$
Catherine~L. Hale,$^{4,6}$
Leah~K. Morabito,$^{2,3}$
Romeel Dav\'e, $^{5,6}$
\newauthor{Matt~J. Jarvis,$^{4,5}$
Robin~H. W. Cook$^{1}$}
\\
$^{1}$International Centre for Radio Astronomy Research, University of Western Australia, 35 Stirling Highway, Crawley, Western Australia 6009, Australia\\
$^{2}$Centre for Extragalactic Astronomy, Department of Physics, Durham University, DH1 3LE, UK\\
$^{3}$Institute for Computational Cosmology, Department of Physics, Durham University, DH1 3LE, UK\\
$^{4}$Astrophysics, University of Oxford, Denys Wilkinson Building, Keble Road, Oxford, OX1 3RH, UK\\
$^{5}$ Department of Physics and Astronomy, University of the Western Cape, Robert Sobukwe Road, 7535, South Africa\\
$^{6}$Institute for Astronomy, University of Edinburgh, Royal Observatory Edinburgh, Blackford Hill, UK\\
}

\date{Accepted XXX. Received YYY; in original form ZZZ}

\pubyear{2015}

\begin{document}
\label{firstpage}
\pagerange{\pageref{firstpage}--\pageref{lastpage}}
\maketitle

\begin{abstract}
We present a qualitative comparison between the host and black hole properties of radio galaxies in the MeerKAT GigaHertz Tiered Extragalactic Exploration~(MIGHTEE) survey with the radio galaxy population in the \simba\ suite of cosmological hydrodynamical simulations. The MIGHTEE data includes a $\sim$1\,deg$^{2}$ pointing of the COSMOS field observed at 1.28\,GHz with the MeerKAT radio telescope and cross-matched with multi-wavelength counterparts to provide classifications of high and low excitation radio galaxies (HERGs and LERGs) along with their corresponding host properties. We compare the properties of the MIGHTEE HERGs and LERGs with that predicted by the \simba\ simulations where HERGs and LERGs are defined as radio galaxies dominated by cold or hot mode accretion respectively. We consider stellar masses $\mstar$, star formation rates SFR, AGN bolometric luminosity $\lbol$, and Eddington fraction $\fedd$, as a function of 1.4\,GHz radio luminosity and redshift. In both MIGHTEE and \simba, the properties of HERGs and LERGs are similar across all properties apart from SFRs due to differences in host cold gas content in \simba. We predict a population of HERGs with low $\fedd$ in \simba\ that are confirmed in the MIGHTEE observations and tied to the faint population at low $z$. The predictions from \simba\ with the MIGHTEE observations describe a regime where our understanding of the radio galaxy dichotomy breaks down, challenging our understanding of the role of AGN accretion and feedback in the faint population of radio galaxies.
\end{abstract}




\section{Introduction}

There is sufficient evidence which shows that supermassive black holes~(SMBHs) exist at the centre of all massive galaxies and co-evolve with their hosts~\citep{Magorrian1998,FerrareseMerritt2000,KH2013}. It is during rapid growth phases where SMBHs can release enormous amounts of energy into the surrounding medium that identifies active galactic nuclei~(AGN). The energy released from the AGN, or AGN feedback, impacts the host galaxy by regulating the star formation within the host galaxy, and can even regulate the cooling of the cluster in which the galaxy resides~\citep{Bower2006, McNamara2007, Fabian2012}.

In some cases, accreting SMBHs can launch relativistic jets observed at radio wavelengths due to the emission of synchrotron radiation from relativistic electrons spiralling around magnetic field lines. These are referred to as radio loud active galactic nuclei~(RLAGN), or radio galaxies, and play an important role in the evolution of galaxies by being responsible for the quenching, and maintaining the quenched status, of massive galaxies~\citep{RJ2004,Best2005a, Best2005b,HardcastleCroston2020}. 
Radio galaxies are typically classified into two different categories, radiative mode (or ``cold mode'', ``radiatively efficient'', ``high excitation'') and jet mode (or ``hot mode'', ``radiatively inefficient'', ``low excitation'')~\citep{BestandHeckman2012, Best2014, HeckmanAndBest2014}. 

Radiative mode radio galaxies commonly exhibit typical features of AGN such as an accretion disc surrounded by a dusty torus. These radio galaxies are typically hosted by massive galaxies that still have some ongoing star formation~\citep{Kauffmann2003}, and accrete efficiently via a geometrically thin, optically thick disc of cold gas~\citep{Shakura1973}. This efficient accretion heats and ionises the surrounding accretion disc and gas which radiates at X-ray, UV and optical wavelengths and, for when there is dust present such as in the torus~\citep{Ramos2017}, radiation is absorbed and re-emitted at infrared wavelengths. This process results in high excitation lines present within their optical spectra. The presence of these lines identify these radiative mode radio galaxies as ``High Excitation Radio Galaxies''~(HERGs). 

Jet mode radio galaxies are thought to accrete via an advection dominated hot medium~\citep{Hardcastle2007} and are hosted by massive early-type galaxies with little to no ongoing star formation~\citep{Best2007}. They emit the bulk of their energy in the form of mechanical jets and do not exhibit typical features of AGN~\citep{MH2007}. That is, accretion disc and torus features, or broad or narrow line regions, etc., are not typically present in jet mode radio galaxies. Jet mode radio galaxies thus lack the presence of or show insignificant high excitation line features in their optical spectra. For this reason they are commonly referred to as ``Low Excitation Radio Galaxies''~(LERGs). 


Previous studies of radio galaxies have shown that HERGs accrete efficiently at Eddington rates of $\fedd>0.01$ while LERGs accrete ineffeciently at $\fedd<0.01$~\citep{BestandHeckman2012,HeckmanAndBest2014,Mingo2014}. This dichotomoy in Eddington rates has been a significant description of the radio galaxy population and has provided the framework for the implementation of black hole feedback prescriptions in cosmological simulations. The majority of simulations have a black hole feedback prescription in the form of a radiative contribution for $\fedd>0.01$ as well as a kinetic contribution at $\fedd<0.01$~\citep{Dubois2012,Dubois2014, Vogelsberger2014, Weinberger2017}.

Further studies showed that, for a moderate range of redshifts, LERGs dominate the radio luminosity function (RLF) at low luminosities while HERG dominate at high luminosities and that the evolution of HERGs and LERGs are distinct. Specifically, HERGs evolve strongly by a constant factor of $\sim$7 up to $z\approx$0.75 for luminosities $\power \sim 10^{25}\WHz$, while LERGs show little to no evolution~\citep{BestandHeckman2012,HeckmanAndBest2014,Pracy2016}. Thus the evolution of the total radio luminosity function (RLF) of AGN is driven by the evolution of these two populations. 

However, \citet{Whittam2018} observed a fainter population of radio galaxies with $10^{21} \la \power \la 10^{27} \WHz$ out to $z=0.7$ and discovered more overlap between the Eddington rates of HERGs and LERGs and that both populations are found across all radio luminosities. An indication that as we probe fainter luminosities, the properties and dichotomy of radio galaxies become less distinct. 

More recently, \citet{Whittam2022} studied the nature of the radio-loud AGN population in the MeerKAT International GHz Tiered Extragalactic Exploration~(MIGHTEE,~\citealt{Jarvis2017},~\citealt{Heywood2022}) survey using the Early Science data in a $\sim$1\,deg$^{2}$ view of the COSMOS field. Radio galaxies were selected out to $z\sim6$ with radio luminosities $10^{20}<\power / \WHz <10^{27}$ and classified as high- and low excitation radio galaxies (HERGs and LERGs) based on whether the radio galaxy is identified as an AGN at other wavelengths apart from radio. Confirming the presence of low accreting HERGs at low luminosities, they found no distinct differences in the host properties of HERGs and LERGs, emphasising the vast overlap in properties of the faint population of radio galaxies. Considering the dominant feedback mechanisms of radio galaxies, the authors additionally found that the minimum stellar mass required to launch radio jets with mechanical luminosity dominating over the radiative output of the black hole is $\mstar \ga 10^{10.5} \msun$ corresponding to a black hole mass of $\mbh \ga 10^{7.5} \msun$ assuming a ~\citet{HR2004} relation of $\mbh \sim 0.0014 \mstar$. 

Sub-grid prescriptions for black hole growth and feedback are vital for cosmological simulations to reproduce properties of the observed galaxy population such as the galaxy stellar mass function, galaxy colour bimodality, the $\mbh-\sigma$ relation, etc~(for a review see \citealt{Somerville2015}). The ways in which different simulations implement black hole growth and AGN feedback can vary significantly. Most simulations prescribe the growth of black holes via a Bondi accretion model~\citep{BondiAndHoyle,Bondi1952,Hoyle1939} describing the accretion via spherical advection of hot gas. This prescription is in line with the proposed accretion mechanism for LERG populations but does not consider the angular momentum transport of cold gas within an accretion disk. Using high resolution galaxy simulations, \citet{HopkinsQuataert2011} proposed a sub-grid prescription that accounted for accretion due to angular momentum transfer from disk instabilities. This ``torque-limited accretion'' prescription was later implemented by \citet{DAA2013,DAA2015,DAA2017} who showed that this model does not require feedback to regulate the growth of the black hole as required in the Bondi model. The torque limited model is now uniquely implemented for cold mode accretion in the \simba\ simulations along with the Bondi model for hot mode accretion~\citep{Dave2019}.

For the feedback from AGN, some simulations prescribe the AGN energy output thermally and isotropically~\citep{Schaye2015,Weinberger2017} while others include both a thermal as well as kinetic and bipolar prescription \citep{Dave2019, Dubois2012, Dubois2014} to account for the morphology of observed radio AGN. While there are multiple prescriptions for feedback, these are typically implemented based on the observed dichotomy of AGN, in that jet mode AGN are inefficiently accreting, and are thus typically activated for Eddington rates $\fedd \lesssim 0.01$.

\citet{Slyz2015} estimated the RLF within Horizon-AGN for both star formation and AGN contributions up to $z=4$, and find consistency with \citet{MS2007} observations. For the AGN contribution to the RLF, the authors select varying Eddington fractions $\fedd$-low and $\fedd$-high to assign radio luminosities estimated from the black hole accretion rate. With appropriate choices of $\fedd$-low and $\fedd$-high, they could broadly reproduce the observed RLF at a range of redshifts. 

Following on from this work, \citet{Thomas2021} used the \simba\ simulations which provide a unique platform to study jet-mode galaxies with the dual hot and cold mode accretion prescription. We were able to identify a population of radio galaxies in \simba\ and classify them into HERGs and LERGs based on their dominant mode of accretion, that is, dominated by cold mode or hot mode accretion respectively. We find a population of inefficiently accreting HERGs with host and environmental properties similar to that of the LERG population,  in particular, at the lower luminosities probed in \simba\, there is significantly more overlap in the host properties of HERGs and LERGs and that the accretion efficiencies of these populations are indistinct, contrary to previous observations.  The predictions from \simba\ however have not been directly compared to observations due to the limited overlap in radio luminosities. It is only now that we can take advantage of the faint radio luminosities, survey area, and depth of the MIGHTEE survey and make direct comparisons with the global properties of the radio galaxy population along with the cosmic evolution of these properties.

This paper is laid out as follows: In \S\ref{sec:mightee} we briefly describe the observations of the MIGHTEE radio galaxy sample and in \S\ref{sec:sims} similarly describe the \simba\ simulations, the radio galaxy criteria therein, and how we match to the MIGHTEE radio galaxy population. In \S\ref{sec:results} we compare the properties of radio galaxies in \simba\ with that of MIGHTEE as a function of cosmic time and radio luminosity. Finally, we discuss these results in \S\ref{sec:discuss} and conclude our findings in \S\ref{sec:conclude}.

\section{MIGHTEE}
\label{sec:mightee}
The MeerKAT International GigaHertz Tiered Extragalactic Exploration~(MIGHTEE, \citealt{Jarvis2017}) continuum Early Science data observed 5 deg$^2$ in the COSMOS and XMM-LSS fields with the MeerKAT L-band receiver covering frequencies 900-1670 MHz~\citep{Heywood2022}. The central 0.86 deg$^2$ in the COSMOS field has been cross-matched to the extensive multi-wavelength data available in the field~\citep{Whittam2024}.

\begin{figure*}
\begin{center}
    \includegraphics[width=0.7\textwidth]{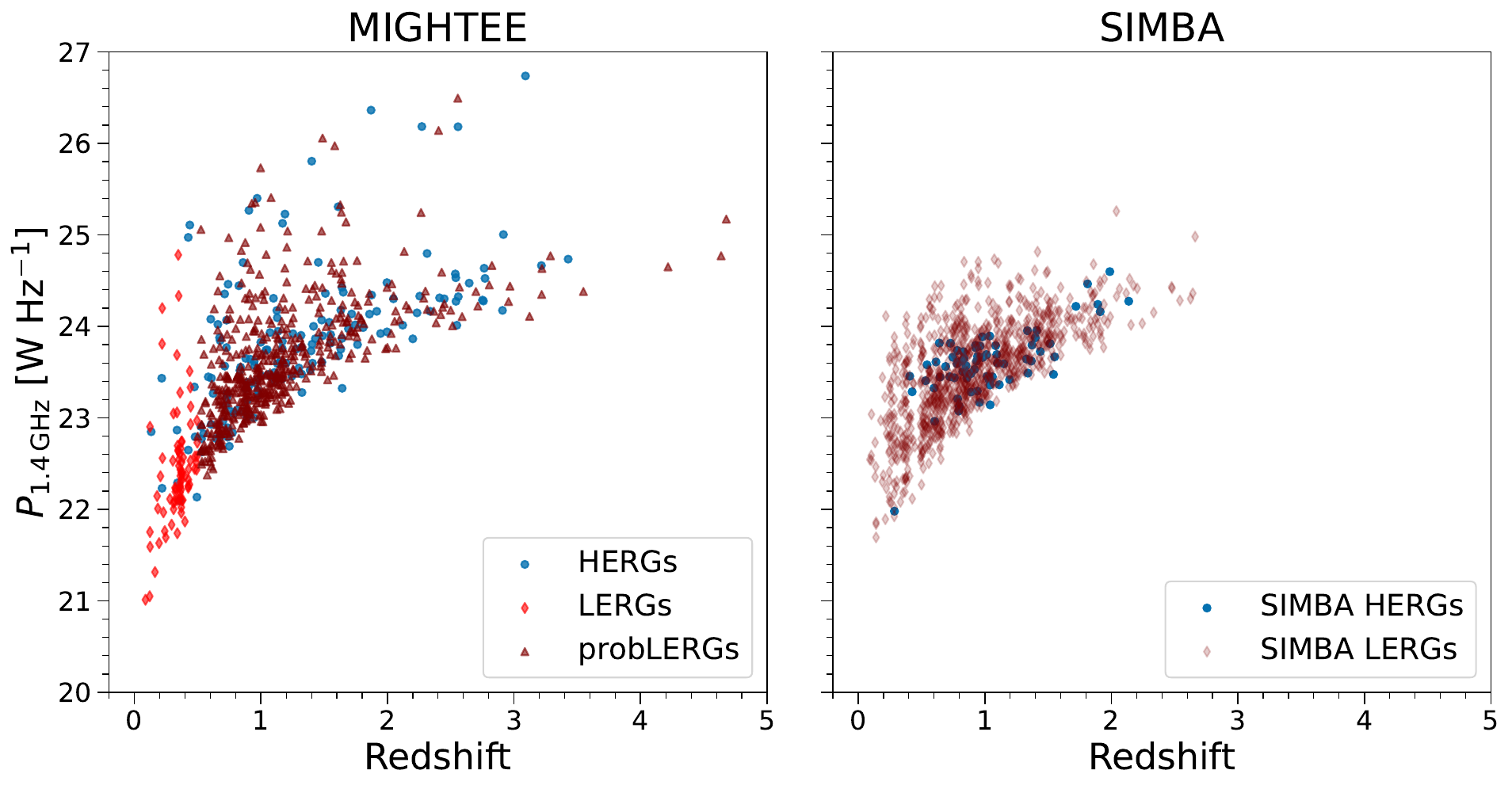}
    \caption{1.4\, GHz radio luminosity $\power$, and redshift of the MIGHTEE Early Science data (left) and \simba\ cosmological simulation lightcone (right). Red diamonds show LERGs, maroon triangles show probable LERGs identified in MIGHTEE, blue circles show HERGs. \simba\ radio galaxies are matched in $\power$ and redshift to that of the MIGHTEE sample for a more directly comparable sample. \simba\ likely misses the highest luminosity sources due to the volume limit of the simulation box and low luminosity prescription for the AGN luminosity.}
    \label{fig:PZ}
\end{center}
\end{figure*}

The cross-matched catalogue consist of 6263 sources with $S_{\rm 1.28GHz} > 20 \muup$Jy. 5223 of these sources have a multi-wavelength counterpart and 1332 are classified as radio excess AGN based on the radio-infrared relation from \citet{Delvecchio2021} in which sources that lie 2$\sigma$ below the radio-infrared relation are classified as radio-excess. Specifically, these sources are selected to have radio emission in excess of what is expected from star formation alone.
These sources are additionally examined for AGN properties at optical, mid-infrared, and X-ray wavelengths using the HST ACS I band image from \citet{Scoville2007}, the \citet{Donley2012} wedge in the mid-infarared colour-colour diagram, and X-ray luminosities $L_{\rm X}>10^{42} erg/s$ respectively. For any radio excess source with an AGN detection in one or more of these criteria, the source is classified as a HERG. If the source is not detected via these criteria, it is identified as a LERG. For LERGs with $z>0.5$, it is possible that the source may not be detected but may have an X-ray luminosity greater than the limit imposed and thus may be misclassified HERGs. For this reason, these LERGs are classified as ``probable'' LERGs (see Figure~\ref{fig:PZ}).

The properties of the MIGHTEE radio galaxy sample are computed via SED fitting of 27 photometric bands from the far infrared to the optical using the SED fitting code {\sc AGNfitter}. These properties, to which we will compare, are stellar mass~($\mstar$), star formation rate~(SFR), AGN luminosity~($L_{\rm AGN}$, here used interchangeably with bolometric luminosity $\lbol$), and mechanical luminosity~($\lmech$) which is derived from the total radio luminosity, $\power$.

To facilitate comparisons between MIGHTEE and other observations and simulations, for a source with flux density $S_{\nu}$ at frequency $\nu$, the radio flux densities and luminosities are scaled to 1.4\,GHz via $S_{\nu} \propto \nu^{-\alpha}$ a spectral index of $\alpha=0.7$, taking into account the variation in effective frequency across the MIGHTEE image. For more details about the classifications and derivation of the properties of the MIGHTEE radio galaxy population, see \citet{Whittam2022}.

In this work we will be combine the LERG and probable LERG samples for a more comprehensive comparison while keeping in mind the potential contamination of HERGs into the LERG populations at $z>0.5$.

\section{Simulations}

\label{sec:sims}
\simba~\citep{Dave2019} is a suite of cosmological hydrodynamic simulations based on the meshless mass code GIZMO~\citep{Hopkins2015}. It is unique in that is employs a two-mode sub-resolution prescription for black hole growth from both cold and hot gas, as well as physically motivated AGN feedback in the form of kinetic winds, jets, and high-energy X-rays. 

For all analysis we use \simba's fiducial $(100\hmpc)^{3}$ box that contains $1024^{3}$ dark matter particles and $1024^{3}$ gas elements which is evolved from $z= 249 \to 0$ assuming a \citet{Planck2016} concordant cosmology with values $\Omega_{m}=0.3$, $\Omega_{\Lambda}=0.7$, $\Omega_{b}= 0.048$, $H_{0}=0.68\kmsmpc$, $\sigma_{8}=0.82$, and $n_{s}=0.97 $.

Halos are identified using a friends-of-friends algorithm applied to all stars, black holes, and gas elements with $n_{H}>$ 0.13 H atoms cm$^{-3}$ which is the density threshold for star formation. Above this threshold, \simba\ stochastically spawns star particles from gas elements with the same mass. Galaxies are resolved at 32 star particle masses which results in a stellar mass resolution of $\mstar = 5.8 \times 10^{8} \msun$ based on the simulation gas mass resolution of $m_{\rm gas} = 1.82 \times 10^{7}\msun$. The dark matter resolution is $m_{\rm DM} = 9.6 \times 10^{7} \msun$.

Galaxies are seeded with a $\mbh = 10^{4} h^{-1}\msun $ black hole at a stellar mass of $\mstar \geq \gamma_{\rm BH} \mbh$ with $\gamma_{\rm BH} = 3 \times 10^{5}$. This threshold is selected such that black hole growth is not suppressed by stellar feedback~\citep{Dubois2015,Bower2017,DAA2017b,Habouzit2017,Hopkins2021}. 
For this black hole seed mass, galaxies are therefore seeded with a black hole at $\mstar \approx 10^{9.5} \msun$. In this work we will generally be considering more massive galaxies that host jet mode AGN feedback.

We use a 1\,deg$^{2}$ \simba\ lightcone~\citep{Lovell2021} to select galaxies from $z=0\to6$. This lightcone is created from \simba\ snapshots which we cross-match with catalogues of pre-computed galaxy properties which are produced from the same snapshots and which are created using the publicly available {\sc YT}-based package {\sc caesar}. Properties include stellar, gas, black hole, and dark matter halo properties for each galaxy. Particle lists for each particle type is included such that additional properties can be computed by the user.

This work uses the radio galaxies with radio luminosities and radio galaxy classifications defined in \citet{Thomas2021}. Because this study follows from previous work, we will briefly summarise the key black hole models in \simba\ relevant for this work and direct the reader to~\citet{Dave2019} and \citet{Thomas2021} for a full description of the \simba\ simulations and nature of the radio galaxy population within \simba\ respectively.

\subsection{Black Hole Growth}
\label{sec:accretion}

\simba\ employs black hole growth via a two-mode accretion model for cold gas ($T<10^{5}K$) via gravitational torque-limited accretion~\citep{DAA2017}, as well as from hot gas ($T>10^{5}K$) via Bondi accretion~\citep{BondiAndHoyle,Bondi1952,Hoyle1939}. 
For accretion from cold gas, the gas inflow rate, $\dot{\rm M}_{\rm Torque}$ is estimated via the gravitational torque model of \citet{HopkinsQuataert2011} which accounts from gravitational instabilities from galactic scales down to the accretion disc surrounding the black hole.
For accretion from hot gas, or via an advection dominated accretion flow~(ADAF), \simba\ uses the Bondi model which is a prescription for black hole growth widely used in galaxy formation simulations \citep[e.g.][]{Springel2005,Dubois2012,Choi2012}. The Bondi model describes the accretion of a black hole with mass $\mbh$ moving at a velocity $v$ relative to a uniform distribution of gas with density $\rho$ and sound speed $c_{s}$.
Both of these modes take place concurrently such that the total accretion onto the black hole is
\begin{equation}
\label{eq:totaccr}
\dot{M}_{\rm BH}=(1-\eta)(\dot{M}_{\rm Bondi}+\dot{M}_{\rm Torque})
\end{equation}
where $\eta=0.1$ is the radiative efficiency of the black hole.




\subsection{AGN Feedback}

The AGN feedback prescription in \simba\ is motivated by the observed dichotomy of accretion rates of AGN~\citep{HeckmanAndBest2014} and their corresponding outflows. \simba\ therefore employs a multi-mode feedback model governed by the instantaneous Eddington ratio of the black hole where the Eddington ratio is computed as:
\begin{equation}
    \label{eq:fedd}
    \fedd= \frac{\eta c \sigma_{T}}{4 \pi G m_{p}} \frac{\dot{M}_{\rm BH}}{\mbh} 
\end{equation}

\simba\ estimates the velocity of outflows of radiatively efficient AGN, also referred to as AGN winds, based on the observed linewidths of X-ray detected AGN in SDSS~\citep{Perna2017a}.  

If $\fedd<0.2$, there is a slow transition to the jet mode where the velocity becomes increasingly higher as $v_{\rm w}\propto$ log\,$0.2/\fedd$, where $v_{\rm w}$ is the wind velocity. The velocity increase is capped at 7000$\kms$ and results in maximum jet speeds at $\fedd\leq0.02$. An additional criterion requiring $\mbh>{\rm M}_{\rm BH,lim}$ is added which is motivated by observations that show that jets only arise in galaxies with $\mbh\gtrsim 10^{8} \msun$~\citep{Mclure2004, Barisic2017}; in \simba\ this is conservatively chosen as ${\rm M}_{\rm BH,lim}=10^{7.5} \msun$. However, in this analysis we consider only galaxies with $\mbh\gtrsim 10^{8} \msun$.

AGN-driven outflows are modelled by stochastically kicking particles near the black holes with velocity $v_w$ with a probability based on the kernel weight and fraction of mass accreted by the black hole and subtracted from the gas particle before ejection~\citep{DAA2017}. 

The momentum outflow choice is based on the inferred energy and momentum inputs from observations of AGN outflows~\citep{Fiore2017,Ishibashi2018}, albeit towards the upper end of the observations, which we found is required in \simba\ in order to enact sufficient feedback in \simba\ to quench galaxies.

\simba\ additionally includes high-energy photon pressure feedback.  The energy input rate due to X-rays emitted by the accretion disc is computed following \citet{Choi2012}, assuming a radiative efficiency $\eta=0.1$.  X-ray feedback is only applied below a galaxy gas fraction threshold of $\fgas<0.2$, and in galaxies with full velocity jets (i.e at $v_{\rm w}\ga 7000\kms$ and $\fedd\leq0.02$). The feedback is applied spherically within $\ro$, providing an outwards kick for star-forming gas and heating for non-starforming gas. This mode is important in understanding the hot gas in galaxy groups and clusters~\citep{Robson2020} and for providing a final evacuation of gas to fully quench galaxies~\citep{Dave2019}, which for instance manifests in green valley galaxy profiles and qualitatively improves agreement with observations~\citep{Appleby2019}.

\subsection{Radio Galaxies}

\subsubsection{Radio luminosity}
We use a 1\,deg$^{2}$ \simba\ lightcone~\citep{Lovell2021} to select galaxies from $z=0\to6$ mimicking the $\sim$1\,deg$^{2}$ MIGHTEE Early Science observations. For each galaxy within the lightcone we estimate radio luminosities from star formation and black hole accretion and classify radio galaxies as in \citet{Thomas2021} by identifying galaxies undergoing full velocity jet feedback. For star formation we estimate the radio luminosity via 
\begin{equation}
    \frac{P_{\rm non-thermal}}{\rm W\ Hz^{-1}} = 5.3\times 10^{21} \left(\frac{\nu}{\rm GHz}\right)^{-0.8} \left(\frac{SFR{\geq 5\msun}}{\msun {\rm yr^{-1}}}\right)
\end{equation}
\begin{equation}
    \frac{P_{\rm thermal}}{\rm W\ Hz^{-1}} = 5.5\times 10^{20} \left(\frac{\nu}{\rm GHz}\right)^{-0.1} \left(\frac{SFR{\geq 5\msun}}{\msun \rm yr^{-1}}\right)
\end{equation}
where $\nu=1.4$~GHz is the observed frequency, and SFR$_{\geq5\msun}$ is the star formation rate in stars more massive than $5\msun$. 
While for black hole accretion we estimate the radio luminosity via the emperical relations in \citet{Kording2008} such that for core emission
\begin{equation}
    \label{eq:rad_all}
    \frac{P_{\rm Rad}}{10^{30}\rm erg\ s^{-1}} = \left( \frac{\mdot }{4\times10^{17}\rm g\ s^{-1}} \right)^{\frac{17}{12}},
\end{equation}
where ${P}_{\rm Rad}\sim \nu { P}_{\nu}$ where $\nu$ is the frequency of the radio emission, in this case $\nu=1.4$\,GHz.  In detail, the core emission likely comes from a region $\sim$tens of pc around the SMBH, which is unresolvable in \simba; we assume the contribution to $\power$ from the remainder of the galaxy (other than from star formation) is small in comparison.

The core emission is applicable for low luminosity AGN and is typically sufficient for \simba\ radio galaxies. This is primarily due to the low jet velocity prescribed in \simba\ along with the volume limit of the simulation which cannot produce rare, bright radio galaxies. In particular, the minimum volume density in a $(100h^{-1} Mpc)^{3}$ simulation box is $3.14 \times 10^{-7}$Mpc$^{-3}$mag$^{-1}$ which corresponds to the radio luminosity function at $\power \approx 10^{25} \WHz$~\citep{MS2007} to which the \simba\ radio galaxy population was fit.
To make up for the lack of such sources and account for the contribution from radio lobes to the radio luminosity function, we select 10\% of HERGs that are central galaxies and that reside in the largest dark matter halos, and add a contribution given by:
\begin{equation}
    \label{eq:rad_hergs}
    \log_{\rm 10}\ P_{\rm 151} \left({\rm W\ Hz^{-1} sr^{-1}}\right) = \log_{\rm 10}\ \mdot  +0.15
\end{equation}
We assume a radio spectral index of $\alpha=0.7$ for a source with flux density $S_{\nu}$ at frequency $\nu$ such that $S_{\nu} \propto \nu^{-\alpha}$ to scale from $P_{\rm151~MHz}$ to $\power$.

\subsubsection{HERGs and LERGs}
In \simba, we separate the radio galaxy population into high and low excitation radio galaxies~(HERGs and LERGs) based on their dominant mode of accretion (i.e. $>50\%$ contribution): HERGs are chosen to occur in SMBHs dominated by gravitational torque limited accretion corresponding to cold mode accretion, while LERGs are those SMBH dominated by Bondi accretion corresponding to hot mode accretion. To account for the stochasticity of the accretion model in \simba\ we compute the average accretion rate over 50~Myr.

Selecting radio galaxies in \simba\ via these methods yields a population of faint, inefficiently accreting HERGs with host and environment properties similar to that of the LERG population~\citep{Thomas2021,Thomas2022}. These results are in contrast to observed dichotomy in the radio galaxy population. However, due to the low luminosities probed in \simba, we have been unable to compare directly to observed populations until the release of the MIGHTEE Early Science data. We detail our efforts to make a fair comparison next.

\subsection{Population Matching}

\begin{figure}
\begin{center}
    \includegraphics[width=0.85\columnwidth]{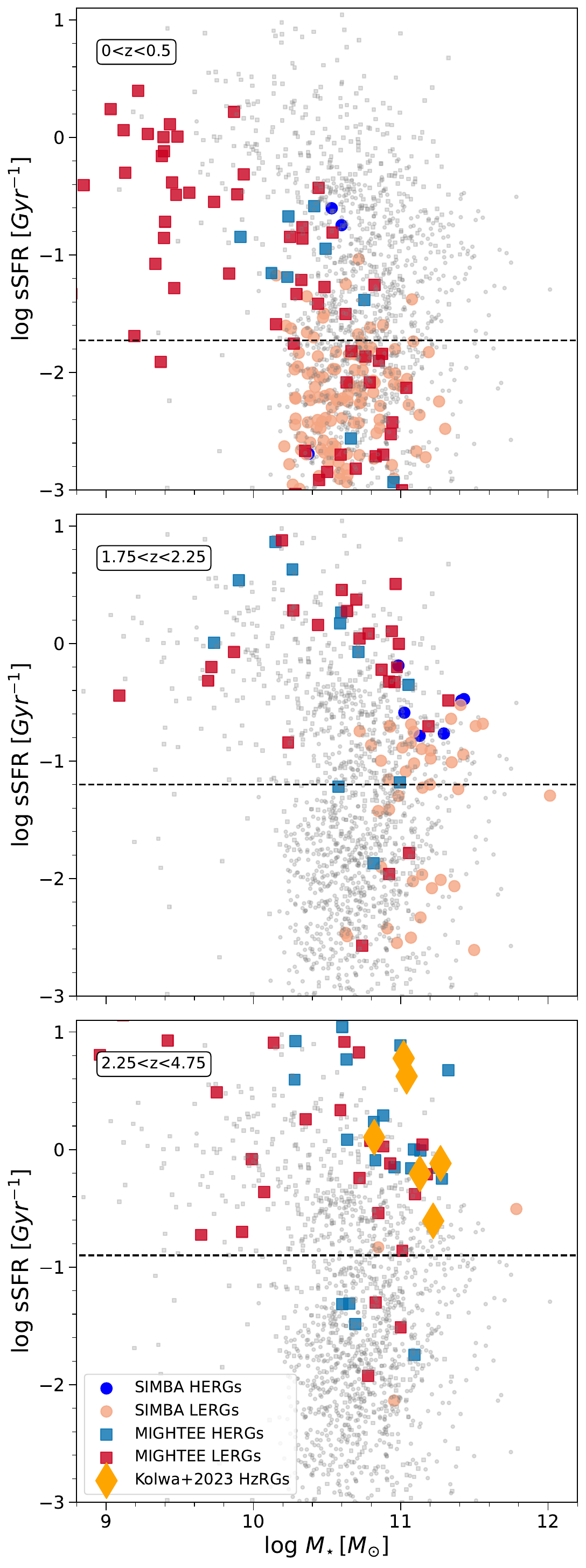}
    \caption{The specific star formation rate~(sSFR) -- stellar mass~($\mstar$) relation for \simba\ and MIGHTEE radio galaxies over 3 redshift bins. From top to bottom, $0<z<0.5$, $1.75<z<2.25$, and $2.25<z<4.75$. Circles indicate the relation in \simba\ for HERGs (dark blue) and LERGs (orange). Squares show the relations for MIGHTEE HERGs (light blue) LERGs (red). Dashed lines indicate the threshold between star forming and quiescence at median redshifts of $z=0.25,2, 3.5$. In the bottom panel, yellow diamonds show the relation for high redshift radio galaxies~(HzRGs) studied by \citet{Kolwa2023}. Grey points throughout are the full sample of radio galaxies in both \simba\ and MIGHTEE to indicate where classifications evolve relative to the full population.}
    \label{fig:mainseq}
\end{center}
\end{figure}

\begin{figure*}
\begin{center}
    \includegraphics[width=\textwidth]{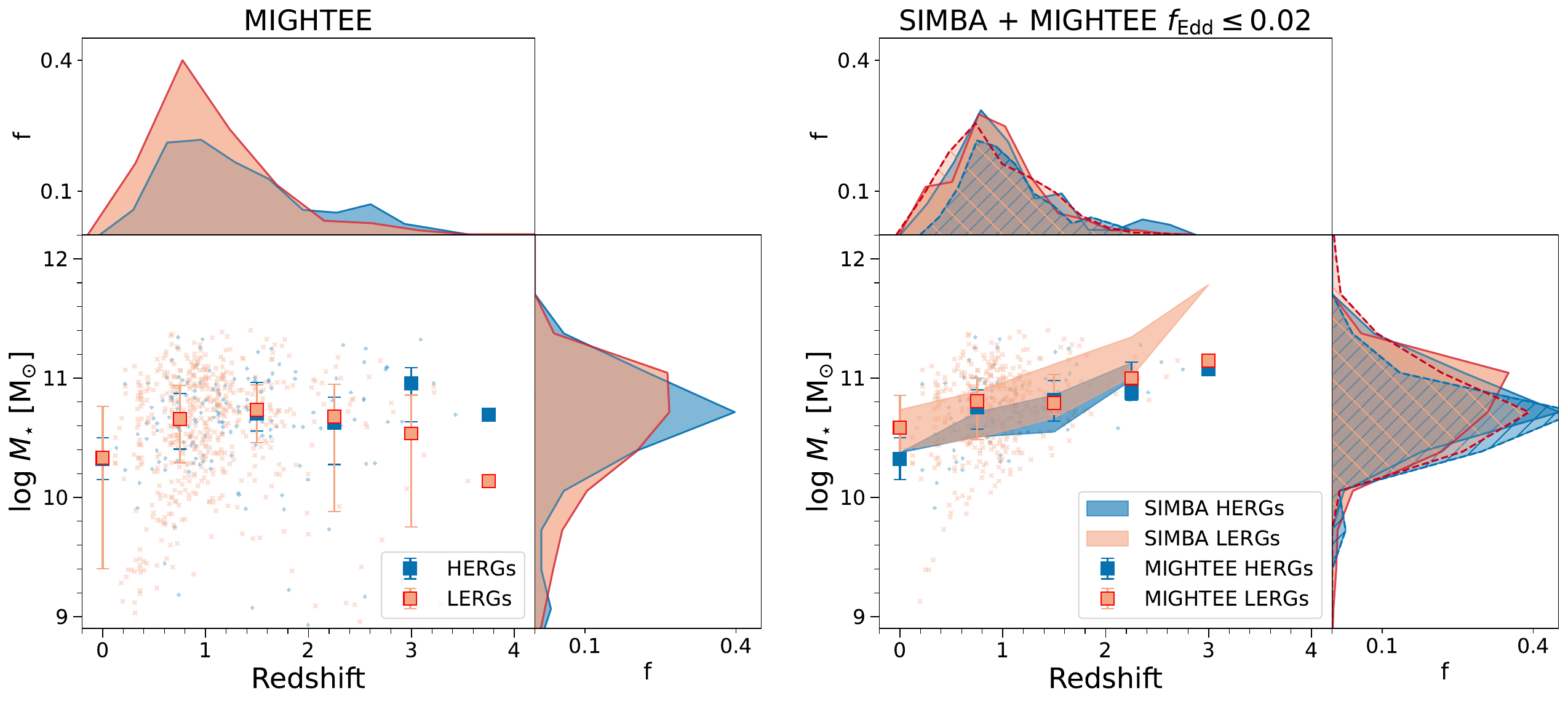}
    \caption{The redshift evolution of the stellar masses~($\mstar$) of the radio galaxy population in the MIGHTEE Early Science data (left, squares), the MIGHTEE Early Science data limited by Eddington rate of $\fedd<0.02$ (right, squares), and the \simba\ cosmological simulations (right, shaded bands). Squares show the median $\mstar$\ per redshift bin with errorbars depicting 1$^{\rm st}$ and 3$^{\rm rd}$ quartiles. Shaded bands similarly show the 1$^{\rm st}$ to 3$^{\rm rd}$ quartile regions for \simba\ galaxies. Additional top sub-panels show the fractional redshift distribution for the radio populations in each sample while the right sub-panels show the fractional stellar mass distributions. Solid line distributions illustrate the MIGHTEE radio galaxies while dashed lines with hatched regions show that for \simba. Blue colours correspond to HERGs while orange colours correspond to LERGs. }
    \label{fig:mstarz}
\end{center}
\end{figure*}

We match the 1\,deg$^{2}$ \simba\ lightcone to the MIGHTEE Early Science data in luminosity and redshift space by interpolating which sources in \simba\ would be detected using the flux-completeness limits of the MIGHTEE survey computed in \citet{Hale2023}. This results in a significant loss of sources with $z>3$. We additionally select all radio galaxies to be radio excess sources that is, lying $0.3$\,dex above the SFR-P$_{\rm 144\,MHz}$ relation of \citet{Best2023} scaled to $\power$ using a spectral index of 0.7. Figure~\ref{fig:PZ} shows the 1.4\,GHz radio luminosity as a function of redshift for the MIGHTEE Early Science data (left) and \simba\ lightcone (right). 
\simba\ is unable to produce high brightness sources and is thus limited to luminosities of $\power < 10^{26}\WHz$ at $z=0$. This is primarily due to the $(100 \hmpc)^{3}$ volume limit of \simba\ and is emphasised for high redshift \simba\ galaxies due to a lower number of jet mode sources but also generally lower black hole accretion rates (which scales directly to the radio luminosity in this case).
This sample of \simba\ radio galaxies will be used throughout the comparison. As we qualitatively compare LERGs in \simba\ with the probable LERG population in MIGHTEE at redshifts $z>0.5$, we will combine the MIGHTEE LERGs and probable LERGs and refer to them as the LERG population throughout this work. We note that the MIGHTEE Early Science data includes sources with $\fedd>0.02$ and thus consider an additional MIGHTEE sample that has $\fedd<0.02$ separately for a more direct comparison with \simba\ jet-mode sources. In \simba, Eddington fractions are computed via equation~\ref{eq:fedd} while for MIGHTEE the Eddington rate is computed as $\fedd = L_{\rm Bol}/ L_{\rm Edd}$ where $L_{\rm Edd} = 1.3 \times 10^{31} \mbh / \msun$~W,  $\mbh \sim 0.0014\mstar$ and $\lbol$ is the AGN contribution estimated from SED fitting with {\sc AGNfitter} i.e the AGN luminosity~\citep{Whittam2022}. LERGs are selected as radio AGN that are not classified as AGN at other wavelengths, that is, where no accretion disk or torus emission is observed. However, LERGs are still passed to {\sc AGNfitter} where they are assigned a significant non-zero AGN luminosity. We thus take caution when deducing results for LERGs based on their $\lbol$.

\section{Radio galaxies in \simba\ and MIGHTEE}
\label{sec:results}

Previous studies of radio galaxies have been limited to sources with high luminosities and reaching redshifts of $z\sim2-3$ \citep{BestandHeckman2012,Williams2018}. There has therefore not been sufficient overlap with the low luminosity sources produced by \simba\ to conduct an appropriate comparison to the observed radio galaxy population. 
Taking advantage of the high sensitivity of the MeerKAT telescope and the $\sim$1\,deg$^{2}$ area of the COSMOS field which has been classified and cross-matched to multi-wavelength data~\citep{Whittam2022,Whittam2024}, we can for the first time directly compare and present the global properties of the population of radio galaxies in the \simba\ simulations with the faint population of radio galaxies in the MIGHTEE Early Science data.

To understand where the host galaxies of the selected radio galaxies in \simba\ and MIGHTEE lie on the main sequence, we show the specific star formation rate~(sSFR) -- stellar mass~($\mstar$) relation for 3 redshift bins in Figure~\ref{fig:mainseq}. The top panel shows this relation for the local population of radio galaxies at $0<z<0.5$. This is also the redshift bin where the MIGHTEE LERGs have secure X-ray non-detections indicating that these are true LERGs in the sense that they are not identified as AGN at wavelengths apart from radio. 
In \simba\ all HERGs lie above the division between star formation and quiescence, while majority of LERGs are quiescent with a few lying within the star forming region. In MIGHTEE most HERGs are star forming, however a vast number of LERGs are also low mass, star forming galaxies. This is a surprising result since past observations of LERGs indicate a dominant high mass, quiescent population~\citep{HeckmanAndBest2014, Whittam2016,BestandHeckman2012}. With increasing redshift, more radio galaxies fall into the star forming region. This is unsurprising due to the higher fractions of cold gas available at earlier cosmic times. There are a small number of quiescent radio galaxies at higher redshifts, however, these are not limited to a specific sub-population. We additionally show the high redshift radio galaxy~(HzRG, $2.9\la z \la 4.5$) sample used to study the gas content in \citet{Kolwa2023}. Both the MIGHTEE and \simba\ HERG populations fall well in line with the gas studies of high redshift radio galaxies.

\subsection{Cosmic evolution of Radio Galaxies}
\label{sec:redshift_evol}
Here we present the black hole and host properties of HERGs and LERGs as a function of redshift.

\subsubsection{Stellar mass}
\label{sec:mstarz}

Figure~\ref{fig:mstarz} shows the redshift evolution of the stellar masses of radio galaxies in the full MIGHTEE Early Science data (left, squares), the MIGHTEE Early Science data limited by Eddington rate of $\fedd<0.02$ (right, squares), and the \simba\ cosmological simulations (right, shaded bands). 

\begin{figure*}
\begin{center}
    \includegraphics[width=\textwidth]{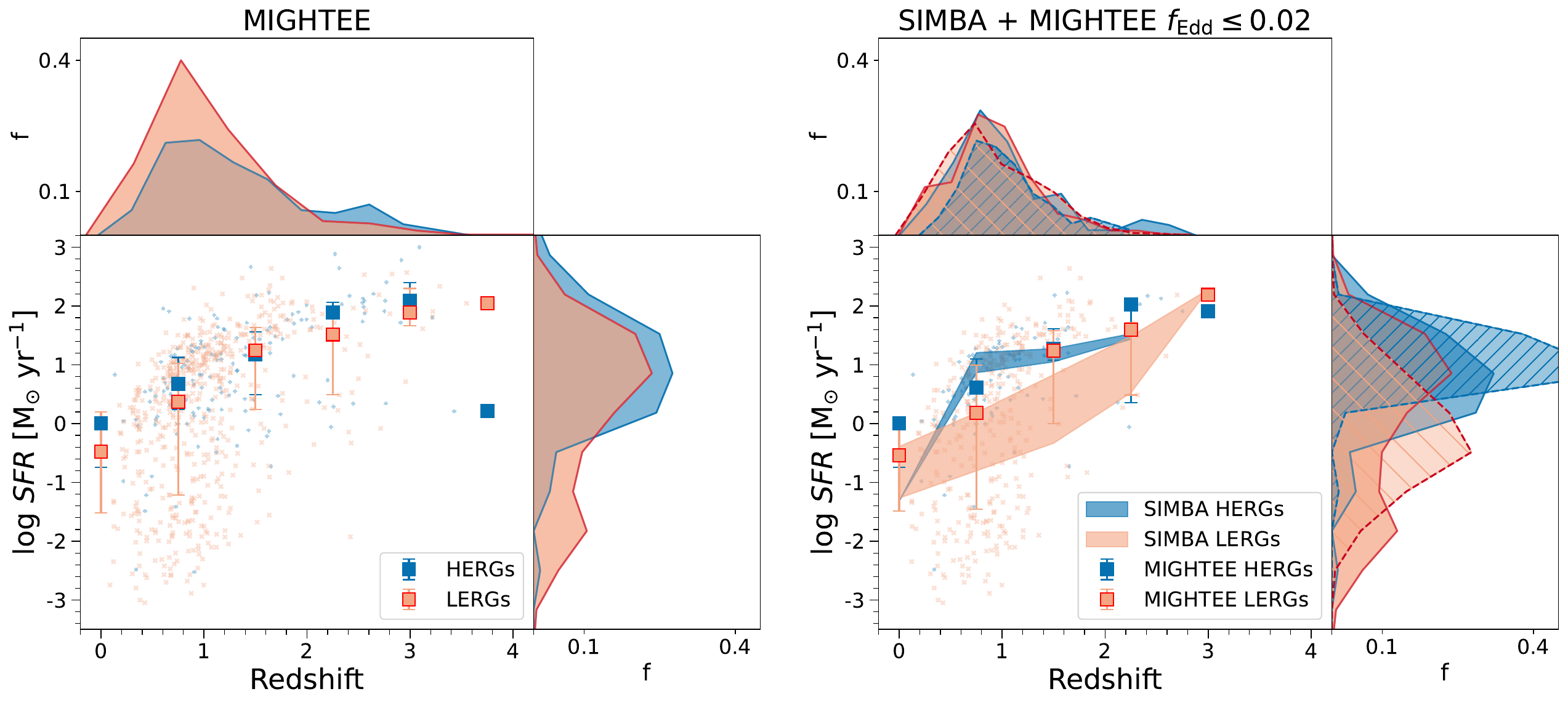}
    \caption{The redshift evolution of the star formations rates (SFRs) of the radio galaxy population in the MIGHTEE Early Science data (left, squares), the MIGHTEE Early Science data limited by Eddington rate of $\fedd<0.02$ (right, squares), and the \simba\ cosmological simulations (right, shaded bands). Squares show the median SFR per redshift bin with errorbars depicting 1$^{\rm st}$ and 3$^{\rm rd}$ quartiles. Shaded bands similarly show the 1$^{\rm st}$ to 3$^{\rm rd}$ quartile regions for \simba\ galaxies. Additional top sub-panels show the fractional redshift distribution for the radio populations in each sample while the right sub-panels show the fractional SFR distributions. Solid line distributions illustrate the MIGHTEE radio galaxies while dashed lines with hatched regions show that for \simba. Blue colours correspond to HERGs while orange colours correspond to LERGs. }
    \label{fig:sfrz}
\end{center}
\end{figure*}

In all three datasets, the median stellar mass increases with redshift. This is tied to the flux limit of the survey as seen in Figure~\ref{fig:PZ} where sources that are much further away at higher redshift are required to be brighter to be detected and thus more massive. This is illustrated by the higher fraction of sources located at $0\leq z \la 2$. We note that \simba\ hosts a number of more massive sources at the highest redshifts than what is seen in the MIGHTEE sample however we attribute this to cosmic variance. Additionally, the MIGHTEE sample hosts more low mass sources at low redshift which is not seen in \simba. This is a consequence of the stellar and black hole mass limits imposed when selecting radio galaxies in \simba\ to fit the observed RLF, that is, $\mstar>10^{9.5}\msun$ and $\mbh>10^{8}\msun$. The sources located at these extreme bins are typically LERGs resulting in low statistical significance when comparing the distributions of LERG stellar masses between the Eddington rate limited MIGHTEE sample with that of \simba\ as illustrated by the low p-value listed in Table \ref{table:kstest}, while the distributions of stellar masses between the HERG populations are similar.

In the full MIGHTEE sample HERGs are hosted by the most massive galaxies beyond $z\sim2.5$, however the overlapping errorbars and low number of sources within these bins makes this statistically uncertain. When limiting the MIGHTEE sample by $\fedd<0.02$, as in the \simba\ sample, we see that this separation disappears and there is no stellar mass preference for radio galaxies at any given redshift bin. Specifically, the limitation on $\fedd$ increases the median $\mstar$ for LERGs in the highest redshift bins. This indicates that the lower $\mstar$ LERGs in the full MIGHTEE sample have higher $\fedd$ than higher $\mstar$ LERGs at high redshift. In general, for both MIGHTEE populations, we see no significant differences between the stellar masses of HERGs and LERGs as a function of redshift nor in total, as illustrated by the fractional distributions on the right side of each panel. Similarly for the \simba\ radio galaxy population, there are no differences in the median and fractional distributions of stellar masses for HERGs and LERGs. MIGHTEE confirms the predictions from \simba\ for the stellar masses when including the faint and low-$\fedd$ population of radio galaxies.

\begin{table}
\begin{center}

\begin{tabular}{|c|c|c|c|c|}
    \hline
      & $\mstar$ & SFR & $\lbol$ & $\fedd$ \\
    \hline
    MIGHTEE-\simba\\HERGs           & 0.0982          & 0.0039         & $\ll 10^{-4}$  & $\ll 10^{-4}$  \\
    MIGHTEE-\simba\\LERGs           & $\ll 10^{-4}$   & $\ll 10^{-4}$  & $\ll 10^{-4}$  & $\ll 10^{-4}$  \\
    \hline
    MIGHTEE         & 0.0761          & $\ll 10^{-4}$  & $\ll 10^{-4}$  & 0.0140        \\
    MIGHTEE-lim     & 0.1555          & $\ll 10^{-4}$  & $\ll 10^{-4}$  & 0.0112      \\
    \simba          &  0.0008         & $\ll 10^{-4}$  & 0.0065        & $\ll 10^{-4}$  \\
    \hline
\end{tabular}
\caption{Kolmogorov-Smirnov~(KS) test p-values for comparisons of distributions of stellar masses $\mstar$, star formation rates SFR, bolometric luminosity $\lbol$, and Eddington rate $\fedd$. In rows 1 and 2 we show the p-values between MIGHTEE-limited and SIMBA HERGs, as well as MIGHTEE-limited and SIMBA LERGs. Additionally, in rows 3-5 we show the p-values between HERGs and LERGs within the full MIGHTEE sample, the MIGHTEE-limited sample, as well as \simba. }
\label{table:kstest}
\end{center}
\end{table}

\subsubsection{Star formation rates}
\label{sec:sfrz}

In the same manner as the previous figure, Figure~\ref{fig:sfrz} shows the redshift evolution of the star formation rates~(SFRs) of radio galaxies in the MIGHTEE, Eddington limited MIGHTEE, and \simba\ samples.

\begin{figure*}
\begin{center}
    \includegraphics[width=\textwidth]{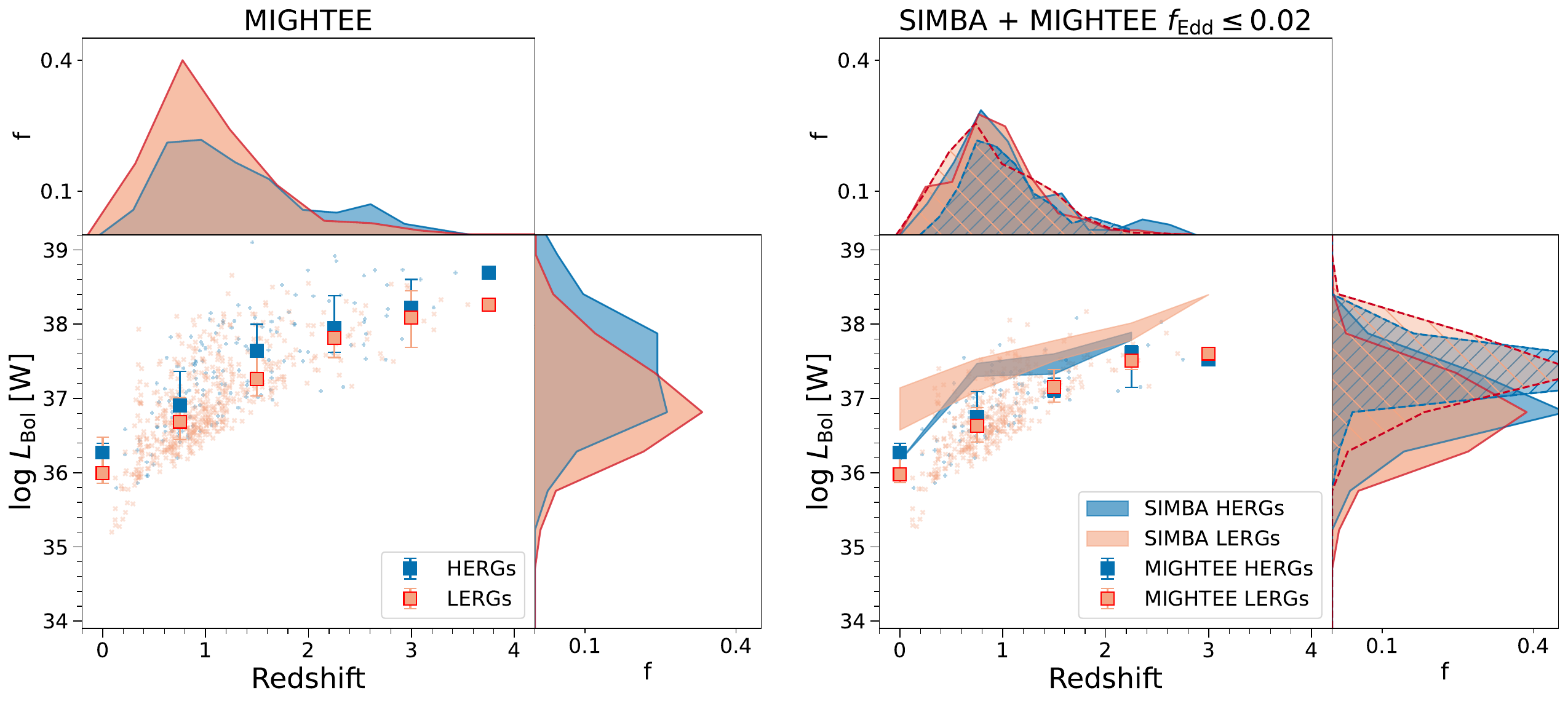}
    \caption{The redshift evolution of the AGN bolometric luminosity ($\lbol$) of the radio galaxy population in the MIGHTEE Early Science data (left, squares), the MIGHTEE Early Science data limited by Eddington rate of $\fedd<0.02$ (right, squares), and the \simba\ cosmological simulations (right, shaded bands). Squares show the median $\lbol$\ per redshift bin with errorbars depicting 1$^{\rm st}$ and 3$^{\rm rd}$ quartiles. Shaded bands similarly show the 1$^{\rm st}$ to 3$^{\rm rd}$ quartile regions for \simba\ galaxies. Additional top sub-panels show the fractional redshift distribution for the radio populations in each sample while the right sub-panels show the fractional $\lbol$ distributions. Solid line distributions illustrate the MIGHTEE radio galaxies while dashed lines with hatched regions show that for \simba. Blue colours correspond to HERGs while orange colours correspond to LERGs. }
    \label{fig:lbolz}
\end{center}
\end{figure*}

In general, the median SFR for radio galaxies in MIGHTEE increases with redshift similarly to that of the total galaxy population \citep[Figure 10]{Whittam2022} due to a combination of flux limitations of high mass sources as well as the increased availability of cold gas at higher redshifts. 
For both the full and Eddington rate-limited MIGHTEE samples at low redshift i.e. $z\la 1.5$, the median SFR for HERGs are higher than that of LERGs, supported by low p-values within the full distributions of each sample. This distinction becomes less clear at higher redshift bins as medians and errorbars overlap. \simba\ shows more significant differences between the SFRs between the distributions of HERGs and LERGs. Specifically in \simba, we tie this to the differences in cold gas fractions seen between the two populations, where HERGs have significantly higher cold gas content than LERGs. Therefore, the dominant mode of black hole accretion taking place within the source is directly linked to the host's gas content. We discuss this in further detail later and in \S\ref{sec:discuss}. Additionally, in MIGHTEE, due to the X-ray undetected data at $z>0.5$, the ``probable LERG'' sample may be contaminated by HERGs that additionally have higher SFRs, leading to the increased SFRs for LERGs and significant overlap with the HERG population.

If HERGs are fuelled by cold mode accretion and LERGs by hot mode accretion, then, perhaps naively, one would expect the host of HERGs to have more cold gas content than that of the LERG populations and that a significant part of the cold gas content will contribute to star formation. The fact, at low redshifts at least, that HERGs have higher SFRs than LERGs indicate that there may be more cold gas in the hosts of HERGs, however the interplay between accretion on the SMBH and star formation is less distinct than what is predicted with \simba. \simba\ divides HERGs from LERGs by the accretion occuring $>50\%$ in the cold mode. While accretion occurs almost completely in the cold or hot mode \citep[Figure 2]{Thomas2021}, it is possible that the threshold for where accretion mode impacts observed properties is rather at a lower ratio of cold to hot accretion. In a given redshift bin, this will classify more sources considered LERGs as HERGs which will decrease the median SFR of the HERG population and match more closely the median HERG SFRs in observations. Similarly, for MIGHTEE, the probable LERG population may be contaminated by HERGs at $z>0.5$ due to limited X-ray observations. If this is the case, removing these sources will decrease the median SFR for LERGs and potentially boost the median SFR for HERGs, resulting in a more distinct dichotomy in SFRs for HERGs and LERGs.

\subsubsection{AGN Bolometric luminosities}
\label{lbol-zs}
Galaxies with high bolometric luminosities are often an indication of enhanced AGN activity. This excess emission occurs at the accretion disc and/or torus in radiative mode AGN where the radiative pressure from the AGN heats the accretion disc and surrounding torus resulting in elevated emission at X-ray to far infrared wavelengths. For jet mode AGN which emit bulk of their energy in the form of kinetic radio jets, we do not expect elevated bolometric luminosities other than what is expected from the host galaxy alone. In other words, we expect that HERGs will exhibit higher bolometric luminosities than LERGs.

For \simba\ we compute the bolometric luminosity contribution from the AGN, or AGN luminosity, as $L_{\rm Bol} = \eta \dot{M}_{\rm BH} c^{2}$ where $\eta=0.1$ is the radiative efficiency. For MIGHTEE, the AGN luminosity is estimated from SED fitting where the host galaxy emission is subtracted from the overall SED and the residual is assumed to be the emission due to the AGN, specifically fitting for emission from the hot dust torus and UV/optical accretion disk components. 

Figure~\ref{fig:lbolz} shows the redshift evolution of the AGN bolometric luminosities of radio galaxies in the full MIGHTEE Early Science data (left, squares), the MIGHTEE Early Science data limited by Eddington rate of $\fedd<0.02$ (right, squares), and the \simba\ cosmological simulations (right, shaded bands).  

\begin{figure*}
\begin{center}
    \includegraphics[width=\textwidth]{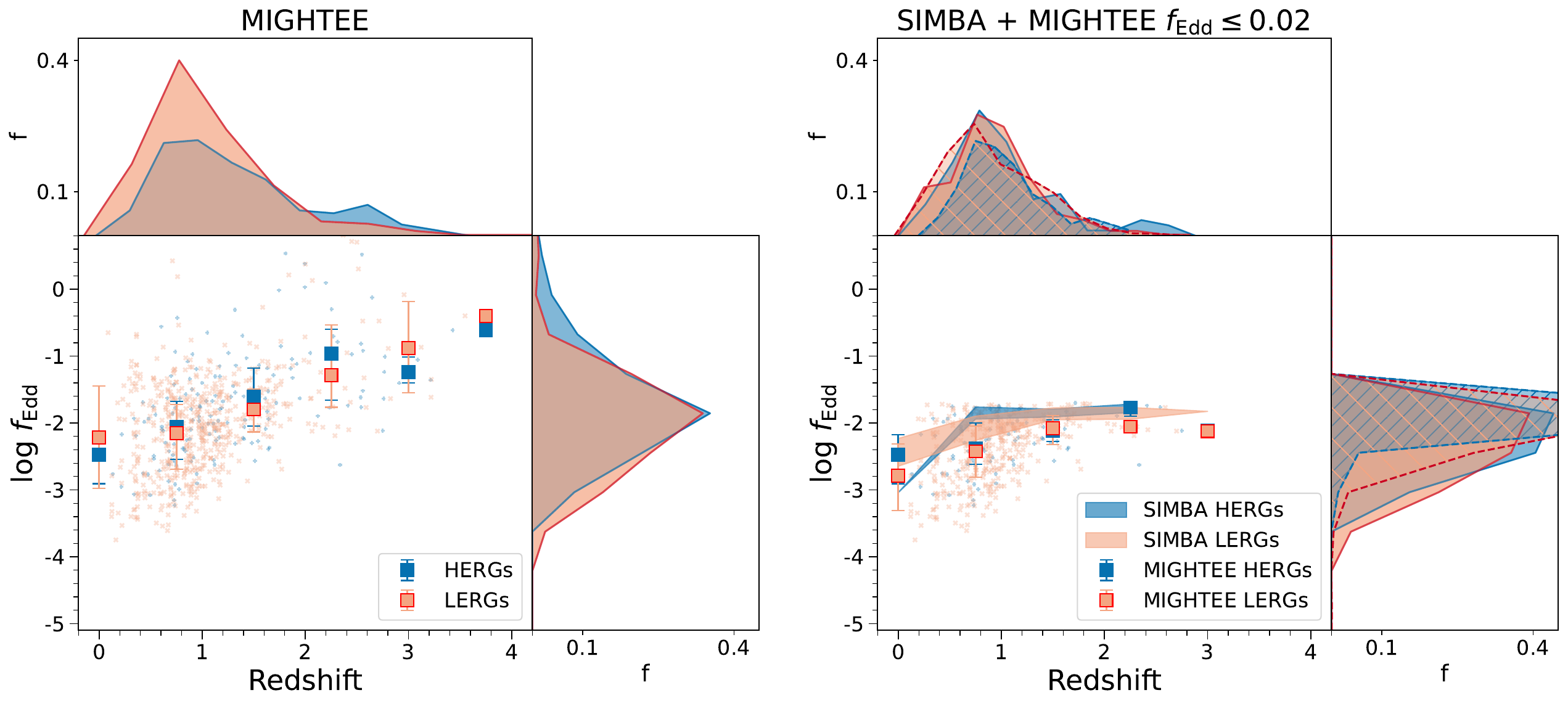}
    \caption{The redshift evolution of the Eddington fraction~(i.e $\fedd = (\lbol+\lmech)/\ledd$) for the radio galaxy population of the MIGHTEE Early Science data (left) and the MIGHTEE Early Science data limited by Eddington rate of $\fedd<0.02$ (right, squares), and the Eddington rate defined as $\fedd = \lbol/\ledd$ in \simba\ cosmological simulations (right, shaded bands). The MIGHTEE data are illustrated by squares which show the median $\fedd$\ per redshift bin with errorbars depicting 1$^{\rm st}$ and 3$^{\rm rd}$ quartiles. Shaded bands similarly show the 1$^{\rm st}$ to 3$^{\rm rd}$ quartile regions for \simba\ galaxies. Additional top and right sub-panels show the respective fractional redshift and $\fedd$ distributions for the radio galaxy populations. Orange colours correspond to LERGs while blue colours correspond to HERGs. Solid line distributions show that of the MIGHTEE sample, while dashed lines with hatched regions show that for \simba. }
    \label{fig:feddz}
\end{center}
\end{figure*}

In the full MIGHTEE sample, HERGs tend to have higher AGN bolometric luminosities across redshift, as expected. While the median $\lbol$ for HERGs at a given redshift is higher than that for the LERGs, this separation is marginal with $\la 0.5$\,dex and within $1\sigma$ uncertainties. Additionally the fractional distributions of bolometric luminosities for HERGs and LERGs are significantly overlapping though still statistically different with LERGs shifted $~0.5$\,dex lower than HERGs. 

For the Eddington limited population of MIGHTEE radio galaxies, at low redshifts where $0\leq z\leq 1$, HERGs have marginally higher $\lbol$ than LERGs. As with the full MIGHTEE sample, the difference between the medians as well as the significantly overlapping errorbars indicates low statistical significance for this result. It is clear however that limiting the Eddington rates of the MIGHTEE sample, removes a large fraction of the high bolometric luminosity sources (dominantly HERGs) and results in fractional distributions that are more similar between populations compared to that of the full MIGHTEE sample.

\simba\ shows a similar increase in $\lbol$ with redshift, however there is more evolution with redshift for LERGs than HERGs along with statistically similar distributions between HERG and LERG populations. However, the values of $\lbol$ for \simba\ are higher than that of MIGHTEE.
Considering the similar values of $\mstar$ for HERGs and LERGs in this region (see right panel Figure~\ref{fig:mstarz}), this indicates that the accretion rates of \simba\ galaxies are higher than that of the MIGHTEE radio galaxy population.

This is a potential consequence of the different ways in which $\lbol$ is measured between observations and simulations. In \simba, all the accretion is encompassed within the bolometric luminosity estimation. Whereas for MIGHTEE, we consider the bolometric luminosities for LERGs to be an upper limit. The reason for this being that LERGs are classified via the lack of AGN signatures at wavelengths apart from emission from mechanical jets at radio wavelengths. This classification implies that there is little to no accretion disk and torus features. LERGs are still passed through the SED fitting code and assigned an AGN bolometric luminosity, contradicting the lack of radiative features assumed during classification of HERGs and LERGs in MIGHTEE. Similarly, while HERGs are classified based on the presence of accretion disk or torus features, they also exhibit mechanical jets and therefore not all the emission due to accretion onto the central SMBH is encompassed by the bolometric luminosity. I.e. for objects with ongoing jet feedback, emission due to accretion processes are in the form of both radiative and mechanical contributions. This caveat is necessary to keep in mind when considering Eddington rates.

A more appropriate comparison of accretion rates between \simba\ and MIGHTEE is to compare $\lbol$ in \simba\ which accounts for all accretion, with the combinations of bolometric and mechanical luminosities in MIGHTEE. We consider this in the next subsection where we define the Eddington rate in MIGHTEE as $\fedd = (\lbol+\lmech)/\ledd$.

\subsubsection{Eddington rates}
\label{sec:fedds}

HERGs and LERGs are typically characterised by a dichotomy in the distribution of Eddington rates. Radiative or High Excitation mode galaxies have been previously thought to dominate at $\fedd>1\%$, and jet or low excitation dominates at $\fedd<1\%$ \citep{BestandHeckman2012, HeckmanAndBest2014,Mingo2014} linking the central SMBH to radiatively efficient and radiatively ineffecient accretion respectively. More recent observations with increased sensitivity however paint a picture which shows more overlap bewtween the two populations \citep{Whittam2018,Whittam2022}.

Figure~\ref{fig:feddz} shows the redshift evolution of the Eddington scaled accretion rate. For \simba\ the Eddington rate is defined as $\fedd = \lbol/\ledd$, and for MIGHTEE this is defined as $\fedd = (\lbol+\lmech)/\ledd$ where $\lmech=7.3 \times 10^{36} (L_{1.4\, \rm GHz}/10^{24} \WHz)^{0.7}$\,W~\citep{Cavagnolo2010}. This additional luminosity is to account for emission in the form of kinetic jets. The full MIGHTEE radio galaxy sample is shown in the left panel while the radio galaxies in MIGHTEE with $\fedd<0.02$ is shown in the right panel overlaid with that of the \simba\ radio galaxy population.

\begin{figure*}
\begin{center}
    \includegraphics[width=\textwidth]{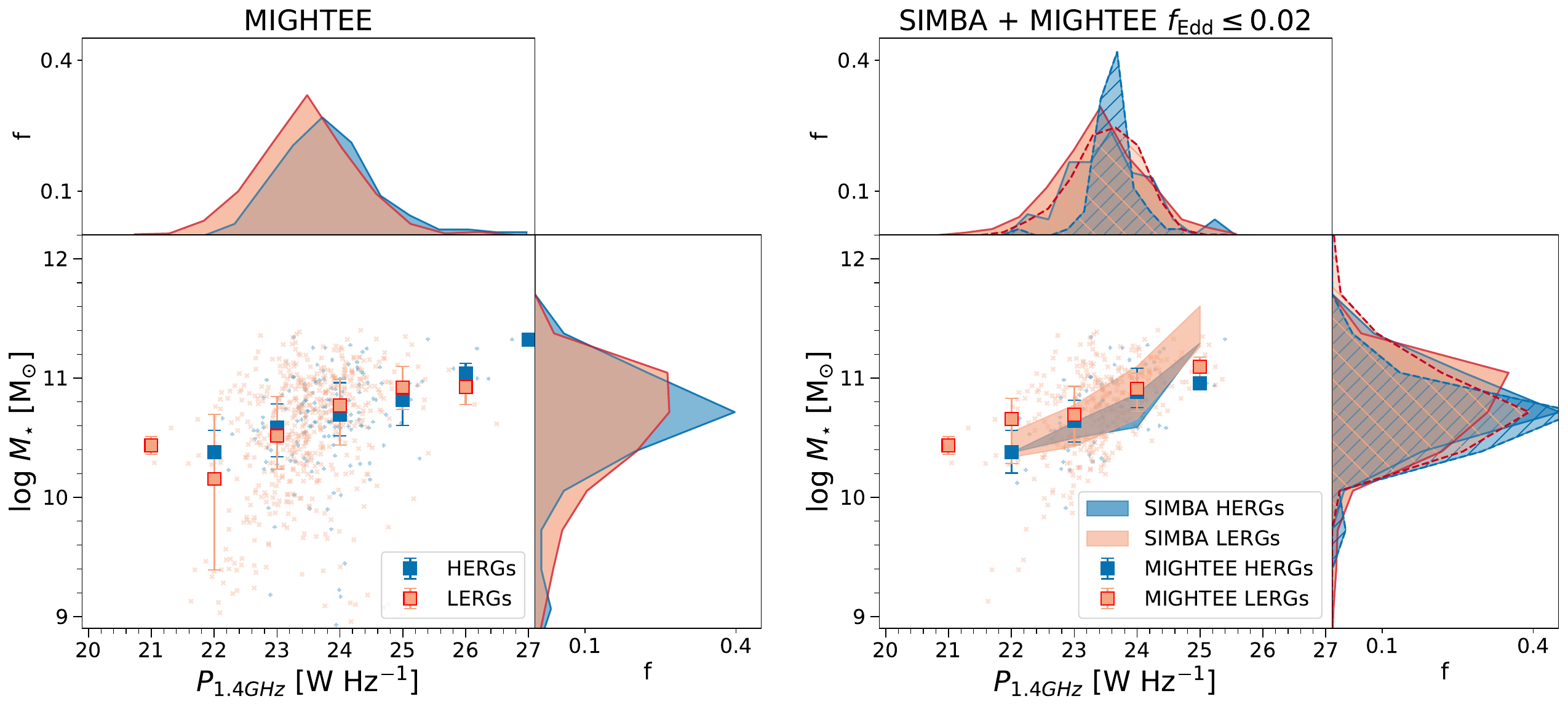}
    \caption{The 1.4\,GHz radio luminosity dependence of the stellar masses~($\mstar$) of the radio population of the MIGHTEE Early Science data (left, squares), the MIGHTEE Early Science data limited by Eddington rate of $\fedd<0.02$ (right, squares), and the \simba\ cosmological simulations (right, shaded bands). Squares show the median $\mstar$\ per luminosity bin with errorbars depicting 1$^{\rm st}$ and 3$^{\rm rd}$ quartiles. Shaded bands similarly show the 1$^{\rm st}$ to 3$^{\rm rd}$ quartile regions for \simba\ galaxies. Additional top sub-panels show the fractional luminosity distribution for the radio populations in each sample while the right sub-panels show the fractional stellar mass distributions. Solid line distributions illustrate the MIGHTEE radio galaxies while dashed lines with hatched regions show that for \simba. Orange colours correspond to LERGs while blue colours correspond to HERGs. }
    \label{fig:mstarp}
\end{center}
\end{figure*}

For the full MIGHTEE sample, there is a significant amount of scatter at a given redshift resulting in a large spread within the distributions of $\fedd$ for both HERGs and LERGs such that the distributions are entirely overlapping. This is supported by having significant p-values between the distributions of $\fedd$ for HERGs and LERGs. Specifically, MIGHTEE confirms the presence of inefficiently accreting HERGs and efficiently accreting LERGs. This result already challenges the previously defined dichotomy in Eddington rates. 
Further limiting the MIGHTEE population to $\fedd<0.02$ we still find overlapping distributions, similarly supported by significant p-values and indicating that the HERG (LERG) populations are not biased toward the high (low) $\fedd$ sources. This is a trend that persists as a function of redshift.

Previously using a $z=0$ snapshot, \citet{Thomas2021} predicted a population of inefficiently accreting HERGs resulting in distributions of Eddington rates between HERGs and LERGs that are indistinguishable. Now, using a 1 deg$^{2}$ lightcone, \simba\ similarly shows an overlapping distribution of Eddington rates between the HERG and LERG population. \simba\ does however span a more narrow distribution toward higher Eddington rates compared to that of the limited MIGHTEE sample, indicating that \simba\ radio galaxies do not include the lowest accreting sources. This could be a consequence of the averaging of the accretion rates due to the time resolution of the simulation. 

We must be cautious when combining $\lbol$ and $\lmech$ as these properties are measured on different timescales in that, the point at which jets were launched is different to the time at which the jet luminosity and $\lbol$ is measured.
In general, \simba\ predicts that the Eddington rates of HERGs and LERGs are indistinguishable across redshift owing to the faint population of radio galaxies that have not been observed before. This prediction is confirmed by the MIGHTEE radio galaxy population.

To briefly summarise the evolution of the properties of the radio galaxy population, \simba\ is able to predict the general trends observed in the MIGHTEE survey. Specifically that, at these low radio luminosities, the stellar masses of HERGs and LERGs are indistinguishable across redshift. The SFRs of HERGs for $z\la2$ are higher than that of LERGs, supporting the hypothesis that the cold gas in the host galaxy is used for both star formation and accretion in HERGs. \simba\ however overestimates the SFRs for HERGs at low redshifts, which can marginally be accounted for by adjusting the threshold for which the accretion is dominated by either torque limited or Bondi accretion instead of the currently chosen 50$\%$. The AGN bolometric luminosities for HERGs are slightly higher than that of LERGs in the full MIGHTEE sample, consistent with the relationship between HERGs and radiative mode AGN. Limiting MIGHTEE to $\fedd<0.02$ removes a large number of the high $\lbol$ sources and disagrees with predictions from \simba\ which, although is limited by $\fedd$, matches better with the bolometric luminosities in the full MIGHTEE sample. This discrepancy is most likely due to the differences in ways $\lbol$ is accounted for between the observations and the simulations, specifically in that all the accretion is accounted for in the case of \simba, however in MIGHTEE we need to consider the emission in the form of mechanical luminosity. Finally, Eddington rates for HERGs and LERGs are indistinguishable when accounting for the sum of AGN and mechanical luminosity, which is in contrast to past observations and the accepted radio galaxy dichotomy. MIGHTEE supports the prediction of a populations of HERGs with $\fedd<0.02$, but also finds a significant contribution of LERGs with $\fedd>0.02$, however we take caution when interpreting Eddington rates for LERGs due to the manner in which $\lbol$ in the SED fitting process is defined, as mentioned in \S\ref{lbol-zs}. Studying the properties of radio galaxies as a function of redshift is a complex notion since it is strongly affected by flux limits of the survey. Next we will consider how the properties of HERGs and LERGs differ instead as a function of radio luminosity.

\begin{figure*}
\begin{center}
    \includegraphics[width=\textwidth]{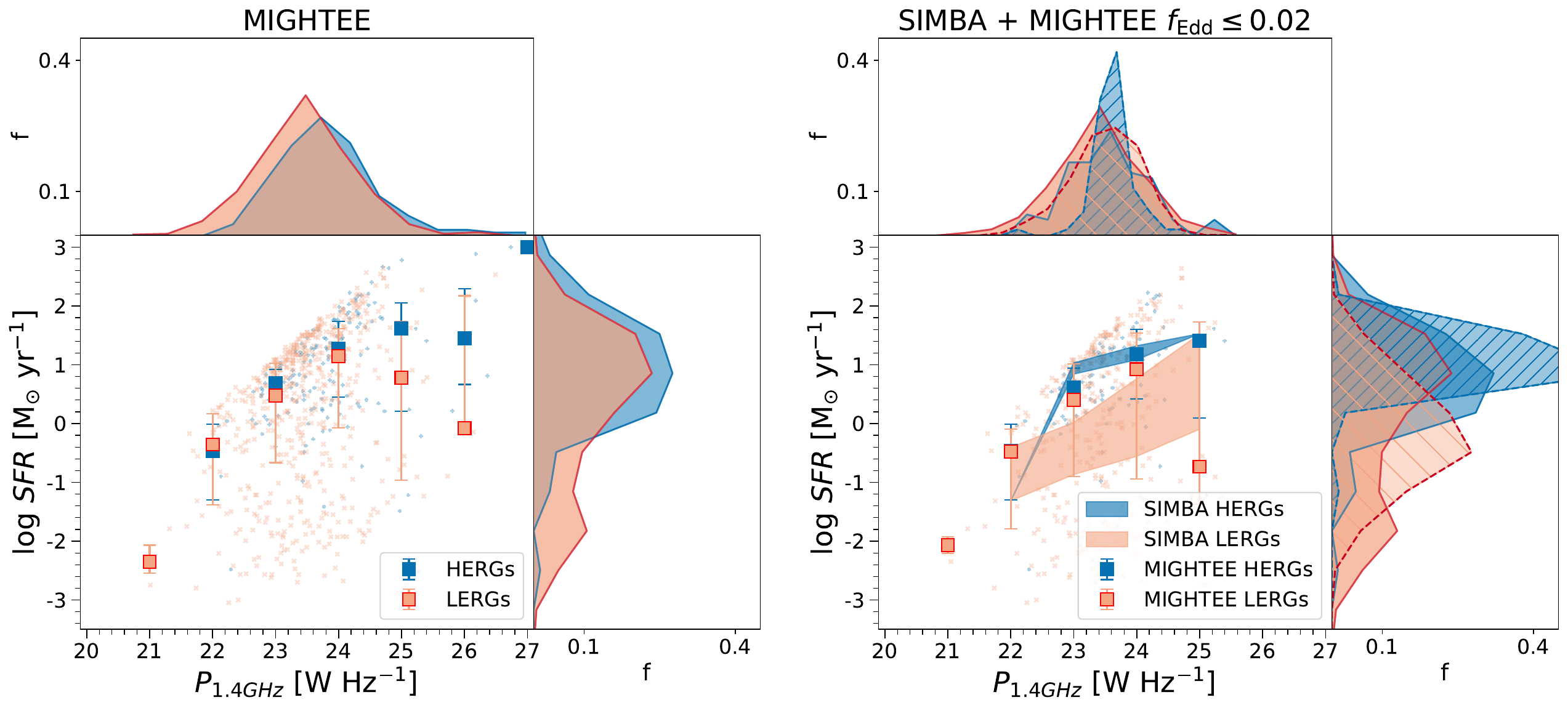}
    \caption{The 1.4\,GHz radio luminosity dependence of the star formation rates~(SFRs) of the radio population of the MIGHTEE Early Science data (left, squares), the MIGHTEE Early Science data limited by Eddington rate of $\fedd<0.02$ (right, squares), and the \simba\ cosmological simulations (right, shaded bands). Squares show the median SFR per luminosity bin with errorbars depicting 1$^{\rm st}$ and 3$^{\rm rd}$ quartiles. Shaded bands similarly show the 1$^{\rm st}$ to 3$^{\rm rd}$ quartile regions for \simba\ galaxies. Additional top sub-panels show the fractional luminosity distribution for the radio populations in each sample while the right sub-panels show the fractional stellar mass distributions. Solid line distributions illustrate the MIGHTEE radio galaxies while dashed lines with hatched regions show that for \simba. Orange colours correspond to LERGs while blue colours correspond to HERGs.}
    \label{fig:sfrp}
\end{center}
\end{figure*}

\subsection{Radio galaxy populations at 1.4\,GHz}
\label{sec:perpower}

Previous studies of the radio galaxy population have been limited by the sensitivity of the observation or survey, that is, limited in both redshift and luminosity. These studies therefore typically probe the brightest sources~\citep{BestandHeckman2012, Whittam2018} where distinctions between the properties of HERGs and LERGs are observed. These constrained populations limit our understanding of the properties of the full population of radio galaxies and whether they are intrinsically different populations. It is therefore crucial to identify whether the same differences between HERGs and LERGs exist as we probe the fainter population of radio galaxies.

\subsubsection{Stellar Mass}
\label{sec:mstarp}
\label{sec:Levolve}

Figure~\ref{fig:mstarp} shows the median $\mstar$ for radio galaxies  as a function of 1.4\,GHz radio luminosity, $\power$. This is shown for the full MIGHTEE Early Science data (left, squares), the MIGHTEE Early Science data limited by Eddington rate of $\fedd<0.02$ (right, squares), and the \simba\ cosmological simulations (right, shaded bands).  

For the full MIGHTEE sample the median stellar masses for both HERGs and LERGs increase with increasing luminosity and with no significant difference between the two populations. There are slight deviations in the highest luminosity bins, however, the number of sources in these bins are not sufficient to make any significant conclusions. 
For the Eddington limited MIGHTEE sample, the low mass end of the LERG $\mstar$ distribution is reduced, indicating that the low Eddington rate LERGs are dominated by the more massive sources, or conversely, the lowest mass LERGs exhibit higher Eddington rates. This is not surprising since the Eddington rate scales with the inverse of the black hole mass which, in the case of MIGHTEE, is estimated from the stellar mass.
As with the full MIGHTEE sample, the number of sources in these bins are too few to make any conclusive statements. The \simba\ radio galaxy sample has a steeper correlation between stellar mass and radio luminosity as well as a much more narrow distribution of radio luminosities. There are no \simba\ HERGs at the faintest luminosities but where HERGs and LERGs overlap there are no differences between in stellar mass for the two population.  
In summary, there are no significant differences between the stellar masses of HERGs and LERGs in both the MIGHTEE and \simba\ populations of radio galaxies as a function of their 1.4\, GHz radio luminosity, even at the highest luminosities.

\begin{figure*}
\begin{center}
    \includegraphics[width=\textwidth]{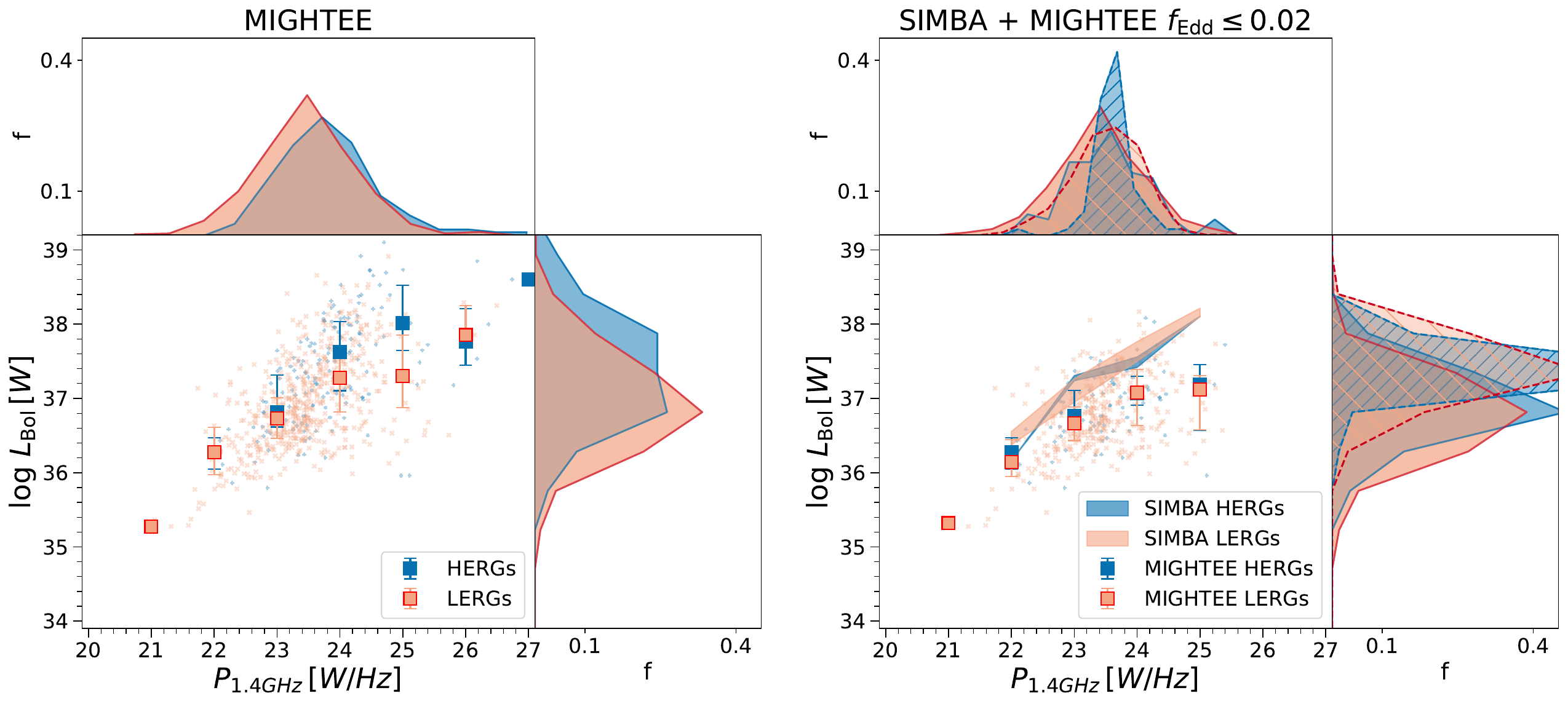}
    \caption{The 1.4\,GHz radio luminosity dependence of the bolometric luminosity~($\lbol$) of the radio galaxy population in the MIGHTEE Early Science data (left, squares), the MIGHTEE Early Science data limited by Eddington rate of $\fedd<0.02$ (right, squares), and the \simba\ cosmological simulations (right, shaded bands). Squares show the median $\lbol$\ per luminosity bin with errorbars depicting 1$^{\rm st}$ and 3$^{\rm rd}$ quartiles. Shaded bands similarly show the 1$^{\rm st}$ to 3$^{\rm rd}$ quartile regions for \simba\ galaxies. Additional top and right sub-panels respectively show the fractional $\power$ and $\lbol$ distributions for each radio galaxy population. Solid line distributions illustrate the MIGHTEE radio galaxies while dashed lines with hatched regions show that for \simba. Orange colours correspond to LERGs while blue colours correspond to HERGs. }
    \label{fig:lbolp}
\end{center}
\end{figure*}

\subsubsection{Star Formation Rates}
\label{sec:msfrp}
Figure~\ref{fig:sfrp} shows the 1.4\,GHz luminosity dependence on the median star formation rates~(SFRs) of radio galaxies in the full MIGHTEE Early Science data (left, squares), the MIGHTEE Early Science data limited by Eddington rate of $\fedd<0.02$ (right, squares), and the \simba\ cosmological simulations (right, shaded bands).  

The SFRs for the full MIGHTEE sample increases with increasing luminosity along the luminosity-SFR relation. Since these are radio excess AGN, all sources have radio emission in excess of what is expected from star formation alone and thus the radio emission in these sources are assumed to be dominated by the central AGN. Specifically, all sources are selected to lie $0.3$\,dex below the infrared -- radio correlation relation of~\citet{Delvecchio2021} for MIGHTEE and 0.3 above the SFR-P$_{\rm 144\,MHz}$ relation of \citep{Best2023} scaled to $\power$ by a spectral index of 0.7 for \simba. This limit is illustrated by the upper envelope on both panels. 
While LERGs dominate at low SFRs, there is still a vast overlap in the distributions of SFRs. At the highest radio luminosities, HERGs have higher star formation rates than equally bright LERGs, however, LERGs still show significant star formation in their hosts - a trend that persists when considering the Eddington limited MIGHTEE sample. There is a downturn in the SFR-$\power$ relation for LERGs at $\sim \power \ga 10^{23.5} \WHz$. This may indicate the threshold at which star formation is switched off and LERGs become quenched by AGN feedback. Considering the full MIGHTEE sample, HERGs do not show this trend clearly, apart from one data point at $\power \sim 10^{25.7} \WHz$. However, when considering the Eddington limited sample, we see a flattening SFR at high luminosities for HERGs. 
The trend seen in HERGs therefore indicate that the accretion and star formation processes for efficient cold mode accretors take place concurrently, while for inefficient cold mode accretion, star formation is reduced at $\power \sim 10^{24} \WHz$ as jet power becomes more powerful. This indicates that efficient accretion in HERGs are linked directly to star formation and thus the gas content available for star formation within the host galaxy. 

For LERGs, the impact of removing high Eddington rate sources results in lower SFRs at the highest luminosities. If LERGs primarily accrete from a hot gas medium, as in \simba, then there is no contest between black hole accretion and star formation for supply of cold gas at $\power \la 10^{23.5} \WHz$. The downturn and potential quenching in the LERG SFRs therefore indicates the point at which hot gas dominates the host galaxy reducing star formation but maintaining the inefficient growth of the supermassive black hole.

For inefficiently accreting LERGs, additional thermal feedback in the form of X-ray heating is provided when host cold gas fractions are below 20$\%$ which further limits star formation in the host. Observationally, it is still unclear whether LERGs are fuelled directly from the advection of hot gas, cooling of this hot gas onto the black hole accretion disk, or whether sporadic accretion occurs from clumpy pockets of cold gas within a hot medium.

As seen before in Figure~\ref{fig:sfrz}, \simba\ shows more significant differences between the median SFRs of HERGs and LERGs. Since HERGs have significantly more cold gas in their hosts, this is beneficial for both black hole accretion and star formation. Additionally, we see that some LERGs still have significant ongoing star formation over several orders of magnitude. However, when considering the fractional distributions, we see that most LERGs have little to no star formation and the HERG distribution peaks at $\sim 2$ orders of magnitude higher than LERGs. 

LERGs in \simba\ do not show a downturn in the SFRs unlike in the MIGHTEE sample, potentially indicating that the jet feedback in \simba\ is not powerful enough to \textit{instantaneously} quench the host galaxy. It is worth emphasising here that radio galaxies in \simba\ are chosen at time of jet launching while MIGHTEE radio galaxies are observed some time after jets have already been launched. An additional reason may be that jets are not quenching sufficiently at higher redshifts, especially when considering that only the $z=0$ \textit{snapshot} does \simba\ predict a downturn in the SFRs of LERGs at $\power \approx 10^{23} \WHz ${~\citep[Figure 7]{Thomas2021}.

\subsubsection{AGN Bolometric Luminosity}

Figure~\ref{fig:lbolp} shows the median AGN bolometric luminosity, $\lbol$, for radio galaxies as a function of 1.4\,GHz radio luminosity, $\power$. This is shown in the full MIGHTEE Early Science data (left, squares), the MIGHTEE Early Science data limited by Eddington rate of $\fedd<0.02$ (right, squares), and the \simba\ cosmological simulations (right, shaded bands).  
At the highest $\power$, HERGs have higher $\lbol$ than LERGs in the full MIGHTEE sample, consistent with previous observations by \citet{BestandHeckman2012, Whittam2018}. This is consistent with HERGs being radiatively efficient and thus having elevated accretion disk and torus emission. However, as the 1.4\,GHz luminosity decreases there is significant overlap between the two populations. Similarly, when considering the fractional distributions of bolometric luminosities (as seen in the right sub-panel), while LERGs show lower $\lbol$ than HERGs, the distributions are overlapping much more than what would be expected if LERGs are expected to show no evidence of AGN activity apart from the radio.

\begin{figure*}
\begin{center}
    \includegraphics[width=\textwidth]{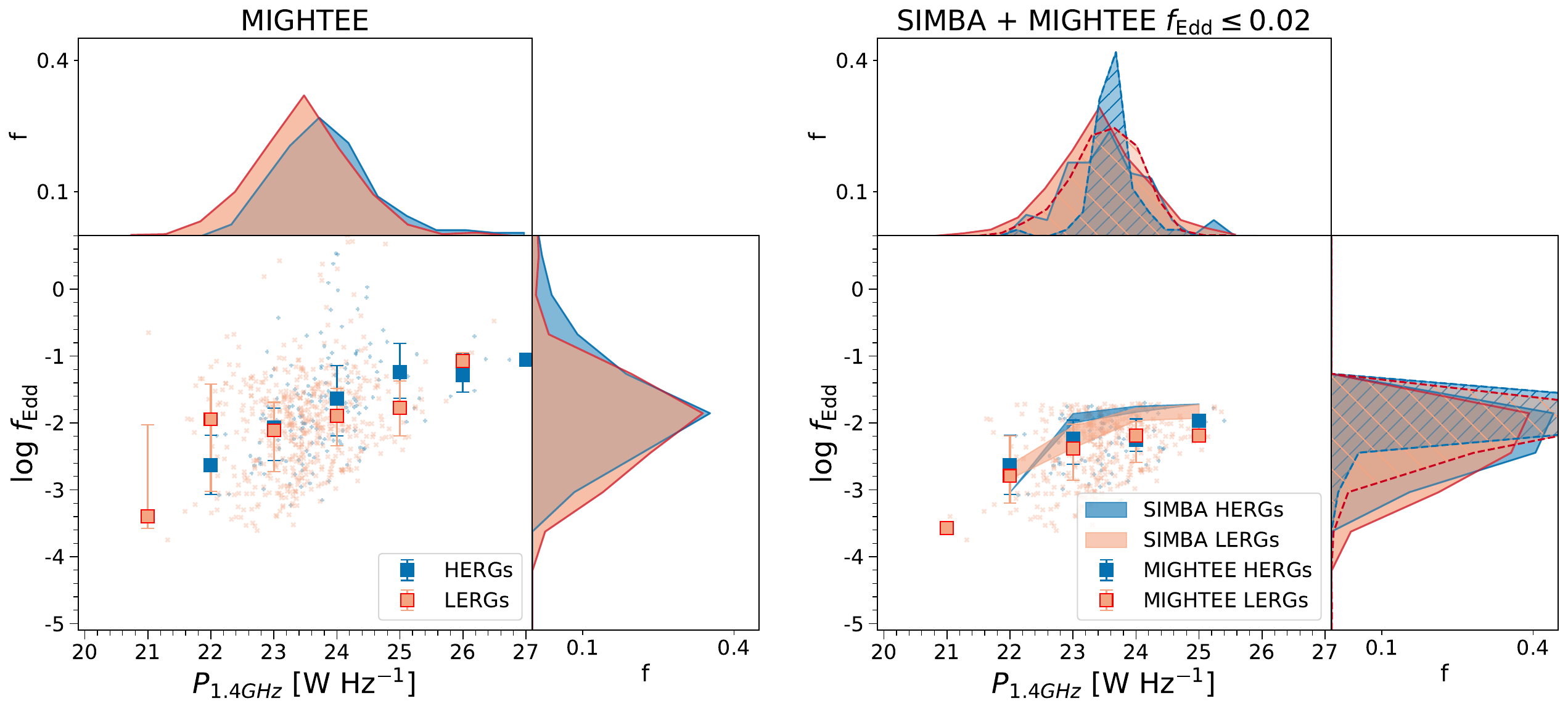}
    \caption{The 1.4\,GHz radio luminosity dependence of the Eddington fraction~($\fedd$) of the radio population of the MIGHTEE Early Science data (left, squares), the MIGHTEE Early Science data limited by Eddington rate of $\fedd<0.02$ (right, squares), and the \simba\ cosmological simulations (right, shaded bands). Squares show the median $\fedd$\ per luminosity bin with errorbars depicting 1$^{\rm st}$ and 3$^{\rm rd}$ quartiles. Shaded bands similarly show the 1$^{\rm st}$ to 3$^{\rm rd}$ quartiles regions for \simba\ galaxies. Additional top sub-panels show the fractional luminosity distribution for the radio populations in each sample while the right sub-panels show the fractional $\fedd$\ distributions. Solid line distributions illustrate the MIGHTEE radio galaxies while dashed lines with hatched regions show that for \simba. Orange colours correspond to LERGs while blue colours correspond to HERGs. }
    \label{fig:feddp}
\end{center}
\end{figure*}

If we limit the MIGHTEE sample to Eddington rates of less than 0.02, we see that the very $\lbol$-bright sources tend to fall away, particularly at high $\power$, and the low $\fedd$ HERGs have lower $\lbol$ much more similar to the LERG counterparts at all $\power$. Note that Eddington rate $\fedd$ directly relates to the AGN bolometric luminosity via $\fedd \propto \lbol/\ledd$ so if the stellar masses, from which black hole masses in MIGHTEE are estimated, are similar across $\power$, limiting by $\fedd$ will result in lower $\lbol$. 

For \simba\ the relation between $\lbol$ and $\power$ is essentially a straight line due to the fact that both properties are estimated from the black hole accretion rate, $\mdot$. We note that, in \simba, the total $\power$ includes emission from star formation but, because we have matched to the observed MIGHTEE sample by selecting only radio excess sources, i.e. those that dominate from AGN emission, star formation processes do not contribute enough to create significant scatter in this relation. 

We remind the reader that LERGs are defined to not have AGN features at wavelengths apart from the radio and so we consider the $\lbol$ in the MIGHTEE sample to be upper limits as the SED fitting code {\sc AGNfitter} assigns them with a significant non-zero $\lbol$. While LERGs may still have accretion disk or torus emission, they may not have enough emission to be identified as an AGN at other wavelengths. The values of the bolometetric luminosities of LERGs in the MIGHTEE sample however are comparable to the HERG population which is a confusing result and suggests that the HERG/LERG classification scheme breaks down in this regime. We further consider the Eddington rate with contributions from both the bolometric and mechanical luminosities next.

\subsubsection{Eddington rates}

Figure~\ref{fig:feddp} shows the luminosity dependence of the median Eddington rate of radio galaxies in the full MIGHTEE Early Science data (left, squares), the MIGHTEE Early Science data limited by Eddington rate of $\fedd<0.02$ (right, squares), and the \simba\ cosmological simulations (right, shaded bands). 

For the full MIGHTEE sample, while LERGs account for the low end of the distribution of Eddington rate sources, the Eddington rates of HERGs and LERGs in a given $\power$ bin are not distinct apart from the bins at $\power \approx 10^{24}$ and $10^{25} \WHz$. This agrees with the notion that it is at the bright end of the radio galaxy population where the dichotomy in Eddington rates are observed. It is difficult to deduce whether this trend holds in the highest luminosity bins due to the low number of sources. However, similarly to the SFRs seen in Figure~\ref{fig:sfrp}, the differences in the Eddington rates of HERGs and LERGs becomes apparent at $\power\approx10^{24} \WHz$, suggesting a threshold at which accretion becomes less efficient in LERGs. Furthermore, tying this to the SFRs indicates the threshold at which cold gas is removed from the host galaxy and id impeded from being accreted for both star formation and black hole accretion processes.

Since $\fedd$ in MIGHTEE includes the mechanical luminosity and since stellar masses in MIGHTEE shows no significant difference between HERGs and LERGs, when comparing to the AGN bolometric luminosities in Figure~\ref{fig:lbolp}, accounting for mechanical luminosity increases the Eddington rates for the LERG population while not significantly impacting that of the HERG population. We note that this is especially true in the highest luminosity bins. This indicates that LERGs are responsible for most of the jet feedback in the bright population of radio galaxies, implying that for low luminosity sources, mechanical feedback is negligable or of the same order of that of the radiative contribution for both the HERG and LERG populations. 

If we consider sources in MIGHTEE that have $\fedd<0.02$ in the right panel, we find populations of both HERGs and LERGs with the same fractional distributions as depicted by the right sub-panel. Specifically, albeit few, inefficiently accreting HERGs can be found across all $\power$. 

As with the AGN bolometric luminosities in \simba, $\fedd$ and $\power$ are both estimated from the black hole accretion rates and therefore there are clear envelopes within the $\fedd-\power$ relation in \simba\ with scatter introduced due to the black hole mass distributions. The \simba\ $\fedd$ matches closely to the median MIGHTEE $\fedd$ as a function of $\power$.

In summary, as a function of $\power$, when considering the distribution of global properties of radio galaxies in MIGHTEE, there are no distinct differences between HERGs and LERGs. However, when approaching luminosities of $\power \ga 10^{24} \WHz$, the two populations begin to diverge. This supports the hypothesis that the dichotomy in the radio galaxy population exists only at the highest radio luminosities and as we approach the faint source population, this dichotomy is reduced. This additionally coincides with the predictions from \simba. \simba\ however shows more significant differences in the star formation rates of HERGs and LERGs. As in \S\ref{sec:sfrz} and shown in Figure~\ref{fig:fgas_z}, this is attributed to the distinct gas fractions in \simba. 

It is worthwhile considering what impact possible contamination within the LERG population has on these results, as well as the redshift evolution of the properties of radio galaxies as a function of their $\power$. While this is beyond the scope of this paper and requires a further in-depth analysis, we show the LERG population in MIGHTEE split into true LERGs~(LERGs with $z\leq0.5$ and thus reliable X-ray non-detections) and ``probable LERGs''~(pLERGs; LERGs with $z>0.5$ and unreliable X-ray non-detections)~\citep{Whittam2022}. This is shown in Figure~\ref{fig:truelergs} of Appendix~\ref{appendixA} where all illustrations are as before with the addition of probable LERGs shown as orange, maroon bordered diamonds. For all four properties analysed in the section apart from stellar mass, there are distinct differences in the ``true'' LERG and ``probable LERG'' populations above $\power \ga 10^{22} \WHz$, in that the probable LERGs match much more closely to the HERG population, and true LERGs follow the trends expected from LERGs in the low redshift universe. This illustrates one of two possibilities. Either there is significant contamination of HERGs within the probable LERG population, or our knowledge of the evolution of accretion and jet-mode feedback above $z=0.5$ is insufficient.

\section{Discussion}
\label{sec:discuss}
We have compared the properties of high- and low- excitation radio galaxies within the \simba\ suite of cosmological simulations with that of the MIGHTEE Early Science Data which covers $\sim$1\,deg$^{2}$ of the COSMOS field. We have found agreements and disagreements when comparing the simulations with the observation that are worth discussing. For where we do find agreement, it is worth noting the difference in the way properties are computed and why these would agree. For example, black hole properties such as black hole accretion rate and black hole masses are followed within the simulations. These properties allow us to directly compute other properties of the radio galaxy such as the bolometric luminosity. These properties are more difficult to derive in observations where spectroscopy is not available. In this case, properties of the black hole and host properties, as well overall classifications of radio galaxies, are estimated from SED fitting of available photometry. This introduces significant uncertainties within the observed results. We consider some of these properties in more detail below.

\subsection{Gas Content in \simba}
\begin{figure}
\begin{center}
    \includegraphics[width=0.4\textwidth]{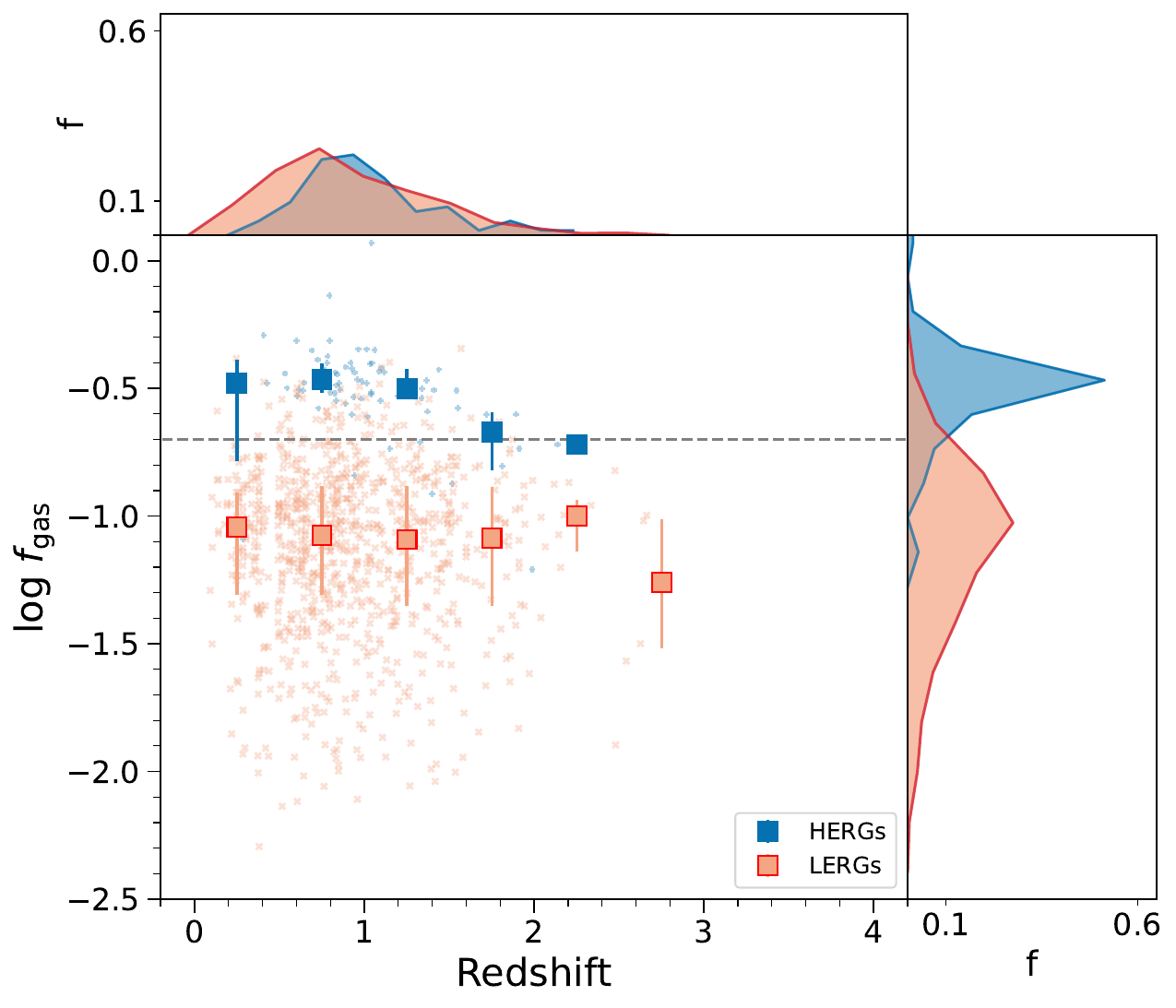}
    \caption{The redshift evolution of the cold gas fractions~($\fgas$) for HERGs (blue) and LERGs (orange) the \simba\ cosmological simulations. Squares show the median $\fgas$ per redshift bin with errorbars depicting $1^{\rm st}$ and $3^{\rm rd}$ quartiles. Additional top sub-panels show the fractional redshift distribution for HERGs and LERGs in blue and orange respectively while the right sub-panels show the fractional $\fgas$ distributions. The grey dashed line shows the $\fgas=20\%$ limit for X-ray feedback in \simba }
    \label{fig:fgas_z}
\end{center}
\end{figure}

Figure \ref{fig:fgas_z} shows the redshift distribution of the fraction of all cold gas in HERGs and LERGs in \simba. 
\simba\ HERGs have $\sim 3\times$ more cold gas than LERGs decreasing slightly toward higher redshifts. The median gas fraction for LERGs stays roughly constant across redshift. Since the stellar masses between the two populations are similar as a function of redshift, the difference in cold gas content results in the significant separation in star formation rates seen in Figure~\ref{fig:sfrz}. The grey dashed line shows $\fgas=20\%$ which is the threshold for the gradual onset of X-ray feedback in \simba. This shows that most LERGs have ongoing X-ray feedback whilst all HERGs do not. There are a number of LERGs that do not have X-ray heating which may account for the contrast seen in star formation rates when tied to the condition for selecting radio galaxies based on their dominant accretion mode.

\subsection{Black hole masses}
For MIGHTEE, black hole masses are estimated from the host stellar mass via a \citet{HR2004} relation of $\mbh \sim 0.0014 \mstar$ which has a scatter of $\sim0.3$\,dex. Within \simba, the black hole mass is tracked throughout the simulation and relates to the stellar mass as $\mbh \sim 0.00125 \mstar$~\citep{Thomas2019}. While using this relation makes no significant difference to results on a global basis, it may be an important detail for single sources or smaller sub-populations.

\subsection{Accretion rates}
In MIGHTEE, bolometric, or AGN, luminosities are estimated by fitting the torus and accretion disk contributions to the SED of each source. For HERGs, which are defined as radio galaxies classified as AGN at other wavelengths, this is a plausible approach. The LERG population is assumed to not form a stable accretion disk and therefore not be classified as an AGN at other wavelengths apart from the radio. However LERGs are assigned significant non-zero luminosity comparable to that of the HERG population, indicating emission associated to accretion and/or torus processes. It is important to question whether these sources are therefore LERGs with boosted UV to Infrared emission or whether the LERG population includes low luminosity HERGs.

This implies that the Eddington rates estimated in MIGHTEE are rather a ratio between AGN bolometric luminosity plus mechanical luminosity and stellar mass which differs to the black hole accretion rate and black hole mass in \simba. The comparison should therefore be treated with caution and used as a guide. While it is difficult to estimate AGN bolometric luminosity from \simba\ in the same way as in MIGHTEE, to determine whether using $\mstar$ impacts our results, we estimated the Eddington rate as defined as before but with black hole masses converted from stellar masses. Doing this shows no significant difference in the respective distributions and results. When converting from stellar mass, the \simba\ radio galaxies Eddington rates are able to increase above $\fedd=0.02$, however the resulting distributions are even more overlapping between HERGs and LERGs in this case. Similarly, as a function of radio luminosity, the relation with black hole accretion rate shows no separation between HERGs and LERGs.

In MIGHTEE, LERGs with lower stellar masses include sources with $\fedd>0.02$, this may be due to higher gas fractions in low mass hosts. MIGHTEE additionally supports the prediction of low $\fedd$ HERGs. Figure~\ref{fig:fedd_dists} shows the fractional distributions of the HERG (blue) and LERG (orange) populations in, clockwise from top left, the full MIGHTEE radio galaxy sample, \simba\ radio galaxies, MIGHTEE radio galaxies limited by $\fedd<0.02$, and MIGHTEE radio galaxies limited by $z<0.75$. The dotted line shows the distribution of $\fedd$ from \citet{BestandHeckman2012}, while the vertical dashed lines show limits at $\fedd=0.01$ and $\fedd=0.02$.
The MIGHTEE sample shows complete overlap between the $\fedd$ distributions of HERGs and LERGs, as we've seen in Figures \ref{fig:feddz} and \ref{fig:feddp}. When constraining the MIGHTEE population to $z<0.75$ we see that the population responsible for this part of the $\fedd$ distribution are faint, low redshift HERGs (see Figure~\ref{fig:PZ}). This challenges our understanding of the interplay between AGN activity and host processes in the faint population of radio galaxies and the role of AGN feedback in these sources. Additionally, at the same redshift limit, we see a bimodal distribution in $\fedd$ for LERGs, this may be due to the contamination of the LERG sample at $z>0.5$, or it may be the contribution of the star forming population of LERGs. This will be investigated further at a later stage.

\begin{figure}
\begin{center}
    \includegraphics[width=0.4\textwidth]{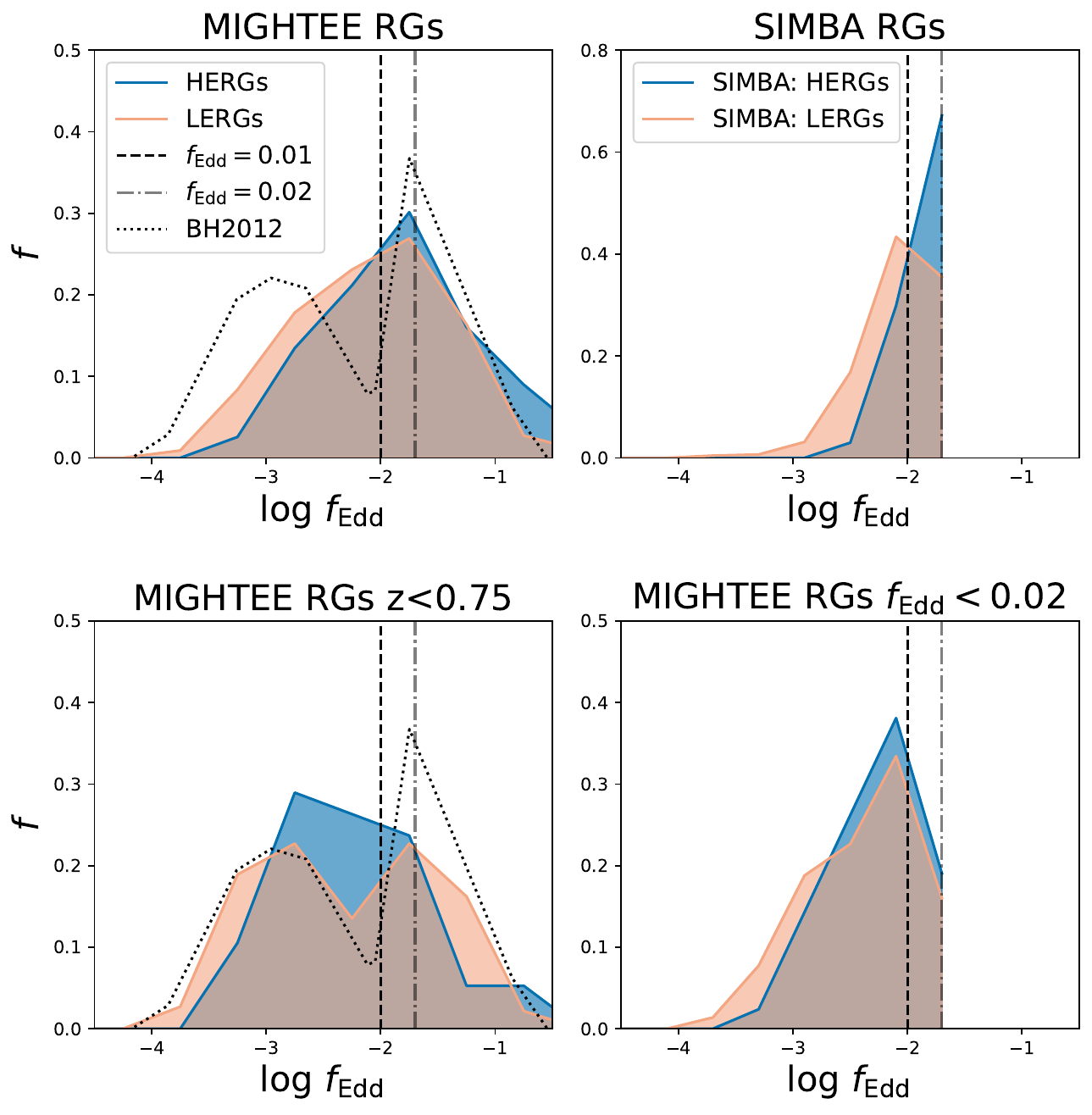}
    \caption{Fractional distributions of Eddington rates $\fedd=\lbol/\ledd$ for HERGs (blue) and LERGs (orange) populations. \textit{Clockwise from top left}: the full MIGHTEE radio galaxy sample, \simba\ radio galaxies, MIGHTEE radio galaxies limited by $\fedd<0.02$, and MIGHTEE radio galaxies limited by $z<0.75$. The dotted line in the top left and bottom left panels show the distribution of $\fedd$ from \citet{BestandHeckman2012}. Vertical dashed lines show limits at $\fedd=0.01$ (black) and $\fedd=0.02$ (grey). MIGHTEE confirms ineffeciently accreting HERG population prevalent in low redshift population of radio galaxies.}
    \label{fig:fedd_dists}
\end{center}
\end{figure}

\subsection{Jet Feedback in cosmological simulations}

A difficulty when making predictions from cosmological simulations for AGN observations, and in particular for radio galaxies, is that feedback prescriptions for jet mode feedback in simulations are based on the observed radio galaxy dichotomy, i.e. that jet mode AGN have $\fedd \la 0.01$ and radiative mode AGN have $\fedd \ga 0.01$. The increasing sensitivity of present radio surveys has shown that this is not the case. Indeed, the existence of powerful radio-loud quasars indicates that jet launching is not unique to inefficient accreting sources. Although these sources exist in the real universe, cosmological simulations typically have too small volumes to produce these types of sources, therefore modelling feedback on the AGN dichotomy is sufficient to produce observed galaxy properties as a whole. 
To add to this, the prescriptions for AGN feedback are also assumed to persist through cosmic time. While little evolution has been seen in the observed radio luminosity function for jet mode radio galaxies, there has been evolution in that of the radiative mode radio galaxies \citep{Williams2018} and there is evolution of the total radio luminosity function in general \citep{Smolcic2017a}. This implies that the impact of jet feedback changes with cosmic time, not only the relative contributions between star formation and AGN, but also the relative contributions from jet and radiative mode radio galaxies, i.e. the radio galaxy dichotomy.   

While there are drawbacks with regard to the implementation of AGN feedback in the current state-of-the-art cosmological simulations, including \simba\, as well as the validity of such a simple estimation of radio luminosities, the prediction of low Eddington rate HERGs is one not expected from the AGN dichotomy, thus the confirmation of this population in MIGHTEE challenges our understanding of radio emission from AGN and the interplay this has with star formation from cold gas content within the host galaxy.

It may be worthwhile considering the role of duty cycles and general timescales under which radio galaxies evolve. That is, the time at which jets are launched from the central SMBH is millions of years before the jets launch their maximum size. The observation of jets emission and that of the central host galaxy therefore does not tell us the conditions under which the jet was initially launched. To make this even more complicated, the timescales at which the accretion of AGN can turn on and off, i.e. the duty cycle, can be on the scale of $\sim$years near the core~\citep{Ulrich1997}, while the duty cycle for jets can be estimated from semi-analytical models to reach a maximum of $\sim 300$\,Myr~\citep{Shabala2020}, indicating that the nature of the SMBH and its Eddington rate, can be changing many times over the time it takes jets to reach their full extent. The reason we observe LERGs to be in quiescent galaxies with low Eddington rates may purely be an indication that the jets have had enough time to inhibit the cooling of gas, thus quenching the host galaxy and inhibiting accretion onto the central SMBH resulting in low Eddington rates. A more appropriate question is perhaps rather what are the conditions at the SMBH at the time of jet launching? Are the hosts of young radio jets HERGs or LERGs and what are their associated properties? These are a series of questions that the Square Kilometre Array Observatory~(SKAO) and it's precursor surveys such as MIGHTEE and the sub-arcsecond resolution counterpart to the Low Frequency Array Two Metre Sky Survey~(LoTSS) observed with the International LOFAR Telescope~(ILT) will begin unfolding~\citep{Morabito2022} by distinguishing star formation and AGN processes based on the morphologies of their emission and probing the fainter populations of radio galaxies not studied before. 

\section{Conclusions}
\label{sec:conclude}
We compare the properties of the radio excess radio galaxy population in \simba\ with the radio galaxy population in the MIGHTEE survey~\citep{Whittam2022}. We compare the stellar masses $\mstar$, star formation rates SFR, and Eddington fractions $\fedd$ as a function of redshift and 1.4\,GHz radio luminosity $\power$. Doing so we find the following results
\begin{itemize}
    \item MIGHTEE and \simba\ agree that HERGs and LERGs do not show significant differences in the distributions of $\mstar$ and $\fedd$. 

    \item HERGs in the MIGHTEE sample have higher SFRs than LERGs, however in \simba\ the separation between the radio galaxy populations is much stronger due to the differences in gas content within the host galaxy. This can be accounted for by adjusting the dominant accretion fraction in \simba. However there may also be contamination of HERGs within the LERG population of MIGHTEE where the X-ray data is insufficiently deep.

    \item  As a function of 1.4\,GHz radio luminosity, the dichotomy in Eddington rates between HERGs and LERGs only becomes apparent at $\power \ga 10^{24} \WHz$.
    
    \item MIGHTEE confirms populations of inefficiently accreting HERGs predicted by \simba. When constraining the population to $z<0.75$, it is clear that this population is prevalent in the faint population of HERGs. 

    \item Even though MIGHTEE LERGs are selected as those without AGN signatures, the population still shows significant $\lbol$ across redshift and $\power$, comparable to that of the HERG population. For radio galaxies with $\fedd<0.02$, it is possible to be inefficiently fuelled by cold mode accretion while maintaining a stable accretion disk and other AGN features. If the LERG population in MIGHTEE is not contaminated by high redshift HERGs with undetected X-ray emission, it is also possible for LERGs to accrete efficiently without exciting AGN features at wavelengths other than the radio.

\end{itemize}
It is important to keep in mind the different ways in which MIGHTEE and \simba\ compute the properties of the radio galaxy population. Using this as a guide, the comparison between the MIGHTEE radio galaxy population and \simba\ highlights the challenges in our understanding of AGN activity in galaxy evolution, specifically, understanding the interplay between AGN accretion and feedback and the interplay with host properties in the faint population of radio galaxies.

\section*{Acknowledgements}
The authors would like to thank Chris Harrison, Martin Hardcastle, and Rohit Kondapally for useful discussions. NLT and LKM acknowledge support from the Medical Research Council [MR/T042842/1]. 
IHW, MJJ and CLH acknowledge generous support from the Hintze Family Charitable Foundation through the Oxford Hintze Centre for Astrophysical Surveys. CLH acknowledges support from the Leverhulme Trust through an Early Career Research Fellowship. MJJ also acknowledges support from a UKRI Frontiers Research Grant [EP/X026639/1], which was selected by the ERC, and the UK Science and Technology Facilities Council [ST/S000488/1]. 
The Simba simulation was run on the DiRAC@Durham facility managed by the Institute for Computational Cosmology on behalf of the STFC DiRAC HPC Facility.  DiRAC is part of the National e-Infrastructure.  For the purpose of open access, the author has applied a Creative Commons Attribution (CC BY) licence to any Author Accepted Manuscript version arising from this submission.

\section*{Data Availability}
The \simba\ data used to derive the findings in this study are publicly available at \url{http://simba.roe.ac.uk} and analysis codes are available upon request.\\

The release of the MIGHTEE Early Science continuum data used for
this work is discussed in \citet{Heywood2022} details of
the data release and how to access the data are provided there. The
cross-matched catalogue and its release is described in the work of
\citet{Whittam2024}. The classification catalogue used in this article is detailed and summarised in \citet{Whittam2022}.


\bibliographystyle{mnras}
\bibliography{simbamightee} 




\appendix
\section{LERG vs probable LERGs}
\label{appendixA}

In the MIGHTEE Early Science continuum cross-matched and classified catalogue, LERGs are separated into LERGs and probable LERGs~\citep{Whittam2022}. This is based on the sensitivity of X-ray data beyond $z=0.5$. Specifically, LERGs (in general) are classified as such based on the lack of AGN features at wavelengths apart from radio. However, above $z=0.5$, the identification of X-ray AGN becomes unreliable as sources could have luminosities above $L_{\rm X} > 10^{42}$erg/s but are not detected. This results in the classification of ``probable LERGs''. It is therefore possible that probable LERGs could be contaminated by the HERG population, or that our knowledge of LERGs above $z>0.5$ is insufficient. Here we consider the properties of LERGs separated into these two sub-classes.

Figure~\ref{fig:truelergs} shows the global properties of radio galaxies as a function of 1.4\,GHz radio luminosity, $\power$, as in \S\ref{sec:perpower}}. That is, we consider stellar mass $\mstar$, star formation rate SFR, AGN bolometric luminosity $\lbol$, and Eddington rate $\fedd$ of MIGHTEE HERGs, LERGs, and probable LERGs in the left column, as well as that limited by $\fedd<0.02$ along with \simba\ HERGs and LERGs in the right column.

Apart from $\mstar$ there are distinct differences in the ``true'' LERG and ``probable LERG'' populations above $\power \ga 10^{22} \WHz$, in that the probable LERGs match much more closely to the HERG population, and true LERGs follow the trends expected from LERGs in the low redshift universe. This illustrates one of two possibilities. Either there is significant contamination of HERGs within the probable LERG population, or our knowledge of the evolution of accretion and jet-mode feedback above $z=0.5$ is insufficient.

\renewcommand{\thefigure}{A\arabic{figure}}

\setcounter{figure}{0}

\begin{figure*}
\begin{center}
    \includegraphics[width=0.65\textwidth]{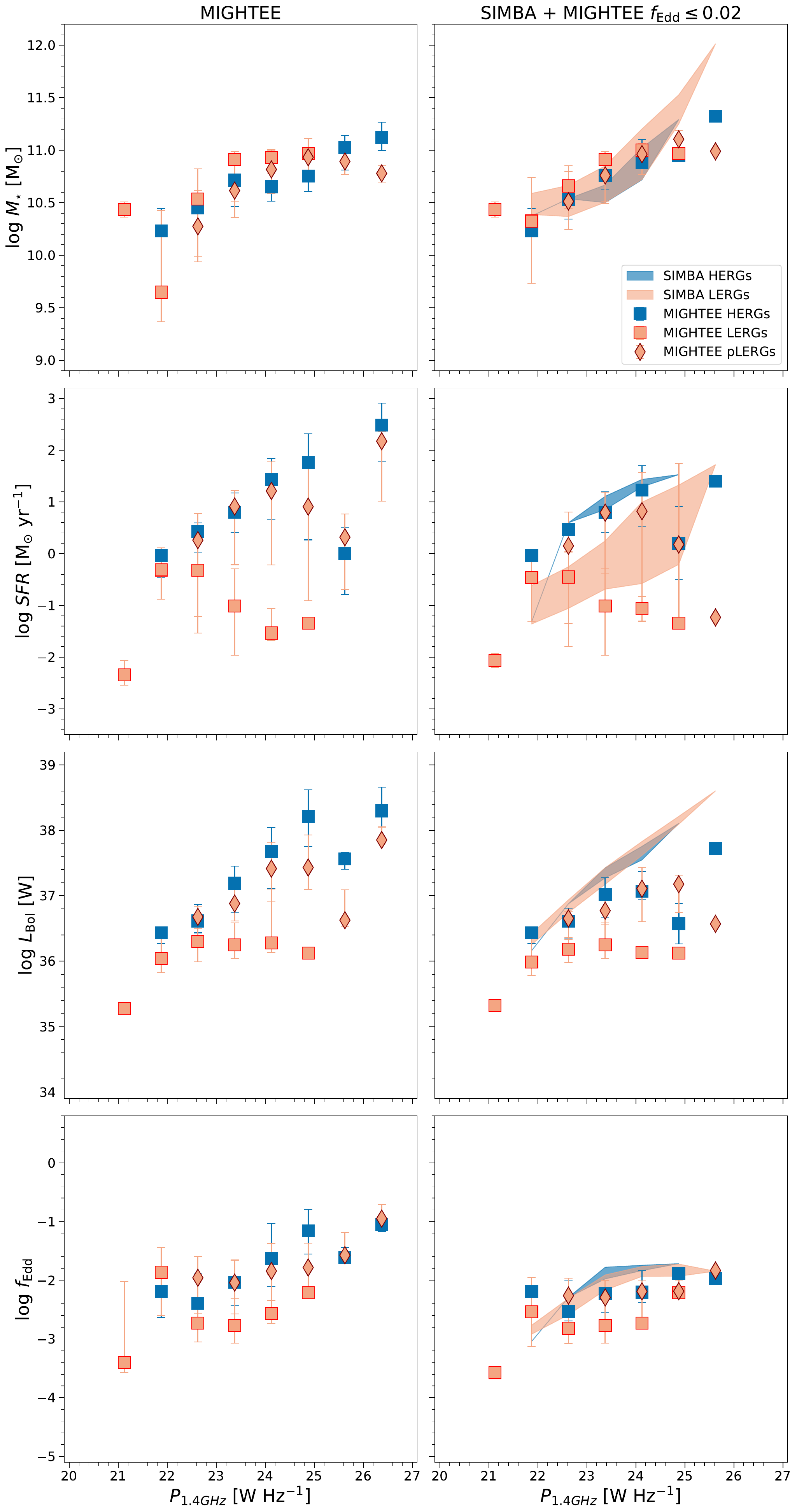}
    \caption{The 1.4\,GHz radio luminosity dependence of the properties the radio population of the full MIGHTEE Early Science data (left column) and MIGHTEE Early Science data limited by Eddington rate of $\fedd<0.02$ (right column) for LERGs (orange squares), probable LERGs (orange diamonds), and HERGs (blue squares) as well the \simba\ cosmological simulations (right column, shaded bands). Points show the median value per luminosity bin with errorbars depicting 1$^{\rm st}$ and 3$^{\rm rd}$ quartiles. Shaded bands similarly show the 1$^{\rm st}$ to 3$^{\rm rd}$ quartiles regions for \simba\ galaxies.}
    \label{fig:truelergs}
\end{center}
\end{figure*}


\label{lastpage}
\end{document}